\pdfoutput=1
\documentclass[PAPER, atlasdraft=false, UKenglish, cernpreprint, texmf, orcidlogo]{atlasdoc}
\usepackage{atlaspackage}
\usepackage{siunitx}
\usepackage{physics}
\usepackage{atlasbiblatex}

\usepackage[particle=true, jetetmiss=true]{atlasphysics}
\graphicspath{{logos/}{figures/}}

\newcommand{\AtlasCoordFootnote}{%
ATLAS uses a right-handed coordinate system with its origin at the nominal interaction point (IP)
in the centre of the detector and the \(z\)-axis along the beam pipe.
The \(x\)-axis points from the IP to the centre of the LHC ring,
and the \(y\)-axis points upwards.
Polar coordinates \((r,\phi)\) are used in the transverse plane,
\(\phi\) being the azimuthal angle around the \(z\)-axis.
The pseudorapidity is defined in terms of the polar angle \(\theta\) as \(\eta = -\ln \tan(\theta/2)\) and is equal to the rapidity
$ y = \frac{1}{2} \ln \left( \frac{E + p_z c}{E - p_z c} \right) $ in the relativistic limit.
Angular distance is measured in units of \(\Delta R \equiv \sqrt{(\Delta y)^{2} + (\Delta\phi)^{2}}\).}

\AtlasTitle{Accuracy versus precision in boosted top tagging with the ATLAS detector}

\AtlasAbstract{%
The identification of top quark decays where the top quark has a large momentum transverse to the beam axis, known as \textit{top tagging}, is a crucial component in many measurements of Standard Model processes and searches for beyond the Standard Model physics at the Large Hadron Collider.
Machine learning techniques have improved the performance of top tagging algorithms, but the size of the systematic uncertainties for all proposed algorithms has not been systematically studied.
This paper presents the performance of several machine learning based top tagging algorithms on a dataset constructed from simulated proton--proton collision events measured with the ATLAS detector at $\sqrt{s} = 13$ \TeV.
The systematic uncertainties associated with these algorithms are estimated through an approximate procedure that is not meant to be used in a physics analysis, but is appropriate for the level of precision required for this study.
The most performant algorithms are found to have the largest uncertainties, motivating the development of methods to reduce these uncertainties without compromising performance.
To enable such efforts in the wider scientific community, the datasets used in this paper are made publicly available.
}

\AtlasRefCode{JETM-2023-06}

\PreprintIdNumber{CERN-EP-2024-159}

\arXivId{2407.20127}

\AtlasJournalRef{JINST 19 (2024) P08018}
\AtlasDOI{10.1088/1748-0221/19/08/P08018}

\hypersetup{pdftitle={ATLAS document},pdfauthor={The ATLAS Collaboration}}

\begin{document}

\maketitle

\tableofcontents

\section{Introduction}
\label{sec:intro}

Collisions at the Large Hadron Collider (LHC)~\cite{Evans:2008zzb} can produce short-lived heavy Standard Model (SM) particles such as the $W$ and $Z$ bosons, the Higgs boson, and the top quark.
These particles often decay to quarks, which later hadronize and are detected as collimated sprays of particles called jets.
In the case that the originating particle's momentum transverse to the beam axis (\pt) is large compared to its mass (i.e. it is \textit{boosted}), the decay products are highly collimated in the laboratory frame and are reconstructed as a single jet.
The task of distinguishing jets resulting from the decays of heavy particles from the much more numerous jets resulting  from light quarks and gluons is known as \textit{boosted jet tagging}.
There is a rich history of boosted jet tagging methods at the LHC~\cite{boost2010, boost2011, boost2012}.
Recently, the adoption of machine-learning-based algorithms for boosted jet tagging has provided large performance improvements~\cite{jss_ml_1, jss_ml_2}.
These algorithms, often called \textit{jet taggers}, have been used in two ways.
The first is to make use of \textit{high-level quantities}, which are observables designed to produce different values when the jet is due to the decay of a heavy particle versus due to light quarks and gluons.
These are always functions of the kinematic properties of the constituent particles (called \textit{jet constituents}) within a jet, which are experimentally reconstructed from inner detector tracks, calorimeter energy deposits, or some non-trivial combination of the two.
A set of high-level quantities is calculated for a given jet, and then used as input to a neural network.
The second approach is to directly use the kinematic properties of the constituent particles as input to a neural network (see Ref. \cite{top_tagging_landscape}).
The information contained in a set of high-level quantities is a subset of the information contained in the kinematic properties of the constituent particles, so the second approach has the potential for higher performance.
However, it requires the use of more complex neural networks as there are a varying and possibly large number of constituent particles within a jet.

The performance of the constituent-based approach has been demonstrated~\cite{ATL-PHYS-PUB-2022-039, ATL-PHYS-PUB-2023-020, ATL-PHYS-PUB-2023-017, ATL-PHYS-PUB-2023-021, ATL-PHYS-PUB-2023-032, CMS-JME-18-002} in the context of highly detailed simulation of the ATLAS~\cite{PERF-2007-01, SOFT-2010-01, SOFT-2022-02} and CMS~\cite{CMS-CMS-00-001} detectors, and used to enhance the sensitivity of several physics analyses at the LHC~\cite{CMS-HIG-21-008, CMS-B2G-22-003, CMS-B2G-21-003}.
However, open questions remain about the relative size of the systematic uncertainties associated with various constituent-based jet tagging algorithms.
The simulated datasets on which ATLAS and CMS train and evaluate ML-based jet tagging algorithms are a useful, but limited approximation of the experimental data gathered by the experiments.
There are important differences that can produce different jet tagging efficiencies between simulated and experimental data.
These differences in efficiency are accounted for through a measuremement of \textit{scale factors}, defined as the ratio of the efficiency in simulated data to the efficiency in experimental data, and its associated uncertainties.
These scale factors are required for interpreting the results of any physics analysis in the context of the SM or any beyond the Standard Model (BSM) physics model, so it is crucial to consider the size of scale-factor uncertainties when comparing jet tagging algorithms.
Scale-factor measurements are a significant bottleneck in the development of new jet tagging algorithms, as they require access to experimental data and must be independently repeated for each algorithm.
As a result they have only been carried out for a few constituent-based taggers~\cite{PERF-2015-04, CMS-PAS-BTV-22-001}, and most studies on constituent-based jet tagging have not considered the uncertainties associated with the application of the tagger to experimental data.

This paper presents the performance of several constituent-based taggers, compared to the performance of a baseline high-level-quantity-based tagger, on a benchmark jet tagging task that involves identifying jets originating from the decay of a boosted top quark.
This task, known as \textit{top tagging}, is a crucial component of many measurements of SM processes that involve the production of top quarks~\cite{TOPQ-2015-10, TOPQ-2018-11, TOPQ-2019-22, TOPQ-2019-28, STDM-2017-34, CMS-TOP-15-015, CMS-TOP-18-013, CMS-TOP-21-014} and searches for BSM physics that contain top quarks in their final state~\cite{CMS-B2G-16-005, EXOT-2019-07, EXOT-2018-48, CMS-B2G-20-002, CMS-B2G-19-004, CMS-B2G-19-003}.
It is used as a benchmark jet tagging task since the three-body decay of the top quark produces jets with a distinctive three-subjet radiation pattern, which can be used to distinguish them from the background of jets originating from light quarks and gluons.

Following the performance comparison, the systematic uncertainties associated with the application of each tagger to experimental data are assessed by applying systematic variations directly to the kinematics of the jet constituents used as inputs to the neural networks.
The resulting variations in tagger efficiency are used to estimate the size of the systematic uncertainties associated with each tagger.
This approach does not require experimental data, and can be easily repeated for an arbitrary tagger by measuring performance on the testing set with the systematic variations. applied, allowing the first comparison of the systematic uncertainties produced by various jet tagging algorithms.
However, as will be discussed, there are many simplifying assumptions made in this approach.
The resulting systematic uncertainties lack the precision needed for use in a physics analysis, but they do provide a useful estimate of the size of the systematic uncertainties associated with each tagger.

The rest of this paper is organized as follows.
The samples of simulated collision events used in this study are described in Section~\ref{sec:mc_simulation}, and the subsequent jet reconstruction and event selections are described in Section~\ref{sec:jets}.
The various jet taggers considered are described in Section~\ref{sec:taggers}, and the performance of the taggers is compared in Section~\ref{sec:performance}.
Section~\ref{sec:uncertainties} describes the procedure used to estimate the systematic uncertainties associated with each tagger, and presents the results.
Finally, conclusions are drawn in Section~\ref{sec:conclusion}.

\section{Monte carlo simulation samples}
\label{sec:mc_simulation}

Simulated proton--proton collisions at $\sqrt{s} = 13$ \TeV using Monte Carlo (MC) methods are used throughout this study.
The nominal samples are generated at leading-order (LO) with \PYTHIA{8}~\cite{pythia8} using the \NNPDF{2.3LO}~\cite{Ball:2012cx} set of parton distribution functions (PDFs) and the A14~\cite{ATL-PHYS-PUB-2014-021} set of tuned parameters.
The effects of pile-up are simulated by overlaying inelastic interactions on top of the underlying hard scattering process.
All simulated samples are passed through a $\GEANT$~\cite{Agostinelli:2002hh}-based simulation of the ATLAS detector.
For more details on the ATLAS detector, see Ref.~\cite{PERF-2007-01}.
Boosted top quarks are selected from simulated events containing the decay of a heavy BSM $Z$ boson ($Z' \rightarrow t\bar{t}$~), with $m_{Z'} = 2$ TeV~\cite{heavy_vector_bosons}.
The cross section of this process is reweighted to produce an approximately flat jet \pt distribution to efficiently populate the kinematic region $[0.35, 5]$ \TeV.
Light-quark and gluon jets are selected from simulated events containing the production of high \pt light quarks and gluons through quantum chromodynamic (QCD) processes.

Additional samples of simulated collisions are utilized to assess the uncertainties from the modeling of the parton shower and hadronization processes as described in Section~\ref{sec:theoretical_uncertainties}.
These uncertainties for boosted top-quark jets are assessed using simulated collision events containing the production of boosted top quarks through SM processes.
Two samples are generated with matrix element calculations performed by the \POWHEGBOX[v2]~\cite{Frixione:2007nw,Nason:2004rx,Frixione:2007vw,Alioli:2010xd} generator at NLO with the \NNPDF[3.0nlo]~\cite{Ball:2014uwa} PDF set and the \hdamp parameter\footnote{The \hdamp parameter is a resummation damping factor and one of the parameters that controls the matching of \POWHEG matrix elements to the parton shower and thus effectively regulates the high-\pT radiation against which the \ttbar system recoils.} set
to 1.5 times the mass of the top quark~\cite{ATL-PHYS-PUB-2016-020}.
For both samples the decays of bottom and charm hadrons were performed by \EVTGEN[1.6.0]~\cite{Lange:2001uf}.
In one sample the parton shower and hadronization is then modeled with \PYTHIA{8}~\cite{pythia8}, and in the other it is modeled with \HERWIG{7}~\cite{herwig1, herwig2}.

The uncertainties from the modeling of the parton shower and hadronization processes in the production of light-quark and gluon jets are assessed using four samples of simulated collisions.
All of these samples are generated with matrix element calculations at LO using the \NNPDF{3.0LO}~\cite{Ball:2014uwa} PDF set.
The parton shower modeling uncertainty is estimated by comparing performance between samples generated with \HERWIG{7}, one produced with the default angular ordered parton shower model, and the other produced with an alternative dipole parton shower model.
The default cluster-based hadronization model is used for both samples~\cite{Winter:2003tt}.
The hadronization model uncertainty is estimated by comparing performance between samples generated with \SHERPA{2.2}~\cite{sherpa}, one produced with the default cluster-based hadronization model~\cite{Winter:2003tt}, and the other produced with the \SHERPA interface to the Lund string fragmentation model of \PYTHIA[6]~\cite{Sjostrand:2006za} and its decay tables.
The default \pt ordered parton shower model is used for both \SHERPA generated samples.

\section{Jet reconstruction, selection, and pre-processing}
\label{sec:jets}

Unified Flow Objects (UFOs)~\cite{JETM-2018-06} are jet clustering input objects that make use of different ATLAS sub-systems in different kinematic ranges\footnote{\AtlasCoordFootnote}.
At low \pt, the inner tracking detector provides exceptional spatial and momentum resolution, so low-\pt charged constituents are reconstructed from tracks using charged Particle Flow objects (PFO)~\cite{PERF-2015-09}.
At high transverse momentum, the tracking detector's momentum resolution degrades but it retains high spatial resolution, and so high-\pt charged constituents are reconstructed using energy measurements from the calorimeters and spatial measurements from the tracking detector using Track Calo Clusters (TCC)~\cite{ATL-PHYS-PUB-2017-015}.
Electrically neutral constituents are reconstructed as neutral PFOs using measurements from the electromagnetic and hadronic calorimeters.
This scheme provides accurate reconstruction of constituent particles across a wide kinematic range.

Both the leading and sub-leading jets in \pt in each event are used in these studies.
Jets are clustered using the \antikt algorithm~\cite{anti-kt} with a radius parameter of $R=1.0$, as implemented in the \Fastjet package~\cite{fastjet}.
The Constituent Subtraction~\cite{constituent_subtraction1, constituent_subtraction2} and Soft-Killer~\cite{soft_killer} (CSSK) algorithms are applied to the neutral UFOs to mitigate contamination from any radiation that comes from pile-up collisions rather than the quarks or gluons that initiated the jet.
Further, the Soft-Drop algorithm~\cite{soft_drop} (SD) is applied with parameters $\beta = 1.0$ and $z_{cut} = 0.1$ to remove soft and wide-angle radiation resulting from pile-up or the underlying event.

Some requirements in the jet selection are placed on the \textit{truth jet}, which is a jet formed from stable particles\footnote{The stable particles are required to have a lifetime greater than 10 ps and muons and neutrinos are excluded as they only leave minimal energy within the calorimeter.} in the simulated event before the detector response is modeled.
All jets are required to have a matched truth jet with $\DeltaR (\text{jet, truth jet}) < 0.75$
The jet itself is required to have a mass of at least 40 GeV, and at least three constituents.
This last requirement is included to ensure the preprocessing scheme described below is well defined.

Jets in the signal sample must satisfy additional requirements which ensure the jet is due to the decay of a top quark, and the decay products of the top quark are fully contained within the jet.
They require that the truth jet be spatially aligned with the observed jet, have a mass greater than 140 \GeV, have a ghost-associated bottom hadron~\cite{ghost_association}, and satisfy a \pT-dependent requirement on the \kt splitting scale $\sqrt{d_{23}}$~\cite{STDM-2012-12}.
For more details on these requirements, see Ref.~\cite{ATL-PHYS-PUB-2021-028}.
All of the requirements are summarized in Table \ref{tab:cuts}.

\renewcommand{\arraystretch}{1.5}
\begin{table}[]
\centering
\caption{A summary of the requirements applied on all of the jets in the simulation samples to produce the training and testing sets. The additional top-quark jet requirements constitute the truth labeling strategy, and are only applied to jets taken from the $Z' \rightarrow \ttbar$ and SM \ttbar samples of simulated events.}
\begin{tabular}{l|l}
\hline
\hline
Jet requirements & Top-quark jet requirements \\ \hline
$\DeltaR (\text{jet, truth jet}) < 0.75$ & $\DeltaR (\text{truth jet, top parton}) < 0.75$ \\
Jet  $|\eta_{\text{true}}| < 2.0$ & Ungroomed truth jet mass $> 140$ GeV \\
Jet  $p_{\text{T},\text{truth}} > 350$ GeV & Number ghost-associated $b$-hadrons $\geq 1$ \\
Number of constituents $\geq 3$ & Truth jet $\sqrt{d_{23}} > \exp (3.3 - \num{6.98e-4} \times \text{truth jet \pT[\GeV]})$ \\
Jet mass $> 40$ GeV &  \\ \hline \hline
\end{tabular}
\label{tab:cuts}
\end{table}

\begin{figure}[t]
\centering
\includegraphics[width=0.6\textwidth]{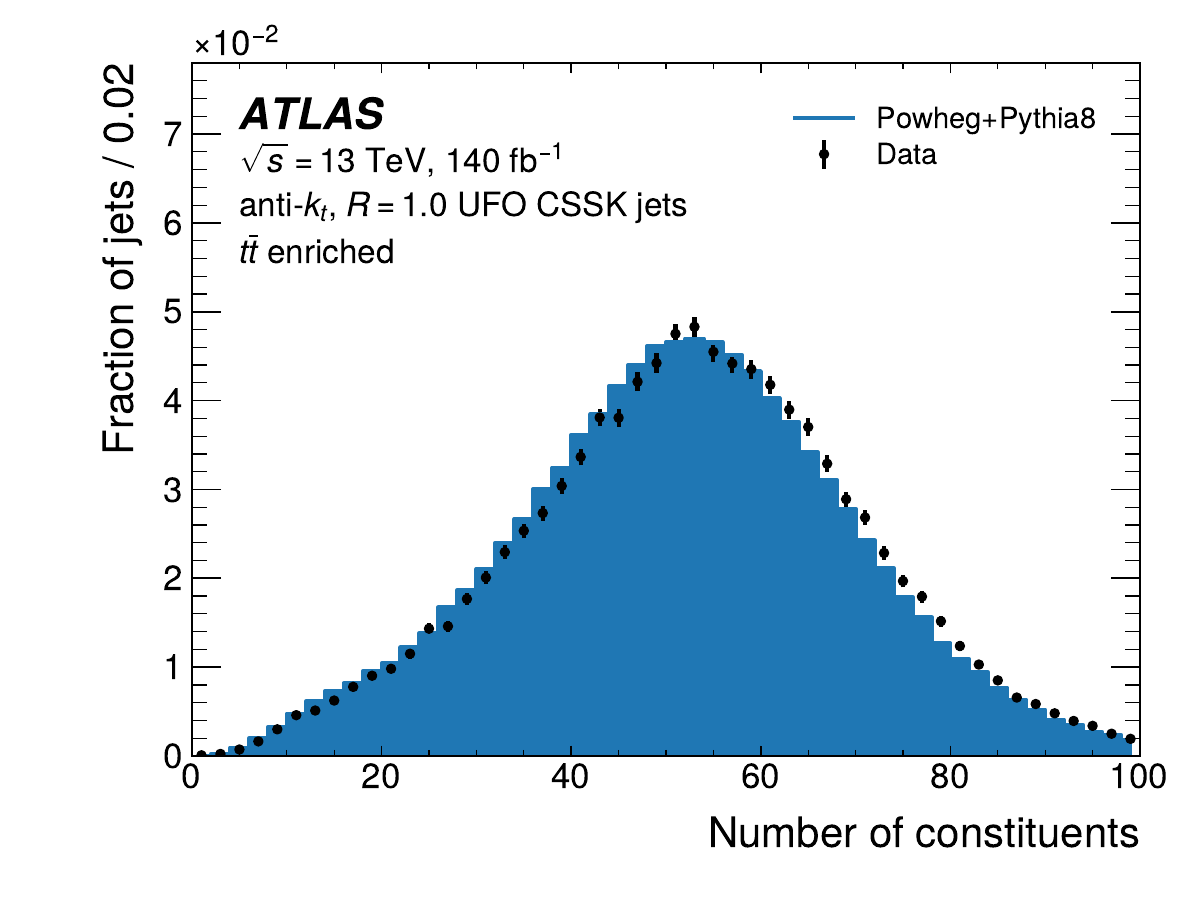}
\caption{
The number of constituents in a sample of jets obtained from the \ttbar enriched region described in Ref.~\cite{JETM-2018-03}. The number of constituents is shown for both the \POWHEGBOX[v2]+\PYTHIA[8] MC sample and experimental data.
}
\label{fig:n_constits}
\end{figure}

The training and performance of ML models can often be improved by applying \textit{pre-processing} to the data to eliminate irrelevant features and capitalize on well-known symmetries.
One set of irrelevant features is the unphysical bumps in the \pt spectrum of the background light-quark and gluon jets, which result from the simulation of QCD multijet events in intervals of jet \pt to allow for efficient generation of high-\pt events.
A physical \pt spectrum is not required for tagger training.
It is only important that the tagger is trained to classify jets from across the desired kinematic range.
To achieve this, the background jet \pt spectrum is re-weighted to match the approximately flat signal jet \pt spectrum.
Additionally, to a good approximation the probability of a jet to be due to a top quark or to a light quark or gluon is invariant under translations of the jet in the $\eta$-$\phi$ plane, and rotations of the jet about the jet axis.
In this study, a pre-processing of the angular coordinates modeled after that used in Ref.~\cite{top_tagging_landscape} is applied to the $\eta$ and $\phi$ coordinates of all jet constituents to remove this approximate traslational and rotational symmetry.
First the coordinates of the jet constituents are shifted such that the highest \pt constituent is located at the origin of the $\eta$-$\phi$ plane.
Then the jet is rotated such that the second highest \pt constituent is located on the negative $\phi$ axis.
Finally if the third highest \pt constituent is located in the negative $\eta$ half-plane, the jet is reflected about the $\phi$ axis to place it in the positive $\eta$ half-plane.
It is also advantageous to pre-process the constituent \pt and energy values to place them on an $\order{1}$ scale.
This is done by taking the logarithm of these values.
Three other constituent-level quantities are calculated and used as inputs to the constituent-based taggers.
The first is the angular distance of the constituent from the jet axis, calculated as

\begin{equation}
R = \sqrt{\eta^2 + \phi^2} ,
\label{eqn:dR}
\end{equation}

where $\eta$ and $\phi$ are taken after the pre-processing.
The second (third) is calculated by dividing the constituent \pt (energy) by the total \pt (energy) in the jet, and then taking the logarithm of this fraction.
All together, seven consituent-level quantities are used as inputs to the constituent-based taggers: the preprocessed $\eta$ and $\phi$ coordinates, the logarithm of the \pt and energy of the constituent, the logarithm of the fraction of the constituent \pt and energy to the total \pt and energy of the jet, and the angular distance of the constituent from the jet axis.

The number of constituents for a selection of jets from the \ttbar enriched region described in Ref.~\cite{JETM-2018-03} is shown in Figure \ref{fig:n_constits} comparing the simulation to a data sample of proton--proton collisions at $\sqrt{s}=13$ \TeV recorded with the ATLAS detector during Run 2 of the LHC and corresponding to an integrated luminosity of 140~\ifb~\cite{DAPR-2021-01,LUCID2}.
Despite the complexity of the ATLAS detector's calorimetry the number of constituents in data is well modeled by simulation.
The jets in this histogram are representative of those entering the signal regions of most analyses targeting boosted top quarks.
Each constituent is characterized by a four-vector, so there is an average of around 200 dimensions used as input to the constituent-based taggers.
To reduce the memory and compute requirements for training the constituent-based taggers, the number of constituents used as input is limited to 80.
Most boosted top jets have fewer than 80 constituents, but those with more are truncated.
The effects of this truncation are mitigated by first sorting the constituents by decreasing \pt, ensuring that only the softest constituents are removed.

\FloatBarrier

\section{Top quark taggers}
\label{sec:taggers}

The top quark taggers considered in this study are described below.
The constituent-based taggers use the four-vectors of the UFOs used to reconstruct the jet as inputs, while the high-level-quantity-based tagger uses the 15 high-level quantities listed in Table~\ref{tab:hlvars} as inputs.
Information useful for identifying heavy-flavor decays, such as the presence of displaced-vertices, is not used as input to any of the taggers.
Instead the taggers are trained to identify the ``3-pronged'' substructure of boosted top-quark jets.
Inclusion of displaced-vertex information in the inputs would likely improve the taggers' performance, but maximizing performance is not the goal of this study and so this is left for future work.
The number of trainable parameters and inference time, defined as the amount of time required to run inference for a batch of 256 jets on an NVIDIA Tesla V100 GPU, are shown for each tagger considered in this study in Table \ref{tab:size_and_speed}.
Many other proposed jet tagging algorithms have shown promising performance in the context of simplified jet reconstruction and detector simulation~\cite{class_party, pelican1, pelican2, pelican3, lorentznet1, lorentznet2}, but these are not considered in this study.
The training, validation, and testing sets consisted of about 9 million, 1 million, and 3.8 million jets respectively, each with equal parts signal and background jets.
The taggers were trained and the hyper-parameters were chosen as described in Ref.~\cite{ATL-PHYS-PUB-2022-039}.

\renewcommand{\arraystretch}{1.5}
\begin{table}[]
\centering
\caption{The number of trainable parameters and inference time for each tagger considered in this study. The inference time is measured using a NVIDIA Tesla V100 GPU.}
\begin{tabular}{c|c|c}
\hline \hline
Tagger  & Number of parameters & Inference time \\ \hline
hlDNN & 133,381 & 3 ms \\
DNN & 876,641 & 3 ms \\
EFN & 959,251 & 4 ms \\
PFN & 754,501 & 3 ms \\
ResNet 50 & 1,499,585 & 20 ms \\
ParticleNet & 764,887 & 143 ms \\ \hline \hline
\end{tabular}
\label{tab:size_and_speed}
\end{table}

\subsection{High-level-quantity baseline}
\label{sec:hldnn}

The high-level quantity densely connected neural network tagger (hlDNN) is trained on the 15 high-level quantities listed in Table~\ref{tab:hlvars}.
This tagger is modeled after Ref. ~\cite{ATL-PHYS-PUB-2021-028}, and serves as the baseline against which the constituent-based taggers are compared.
The network is a standard multi-layer perceptron~\cite{mlp} (MLP).

\begin{table}
\centering
\caption{A listing of the 15 quantities used to train the baseline high-level-quantity-based tagger.}
\begin{tabular}{c|c|c}
\hline \hline
Quantity & Symbols & References \\ \hline
N-subjettiness & $\tau_1$, $\tau_2$, $\tau_3$, $\tau_4$ & \cite{nsubjet} \cite{nsubjet2} \\
$k_t$ Splitting Scales & $\sqrt{d_{12}}$, $\sqrt{d_{23}}$ & \cite{kt_splitting} \\
Generalized Energy Correlation Functions & $ECF_1$, $ECF_2$, $ECF_3$, $C_2$, $D_2$, $L_2$, $L_3$ & \cite{ecf} \cite{ecf_angles} \cite{ecf_power} \\
Minimum Pair-wise Invariant Mass & $Q_w$ & \cite{kt_splitting} \\
Thrust Major & $T_m$ & \cite{thrusts} \\
\hline \hline
\end{tabular}
\label{tab:hlvars}
\end{table}

\subsection{Densely connected neural network}
\label{sec:dnn}

The simplest constituent-based tagger is the densely connected neural network (DNN), which is a multi-layer perceptron operating directly on a vector of the constituent information~\cite{dnn_tagger}.
When there are less than 80 constituents in a jet, this vector contains zero padding which is used as input to the DNN.
The DNN has no mechanism for masking these zero padded inputs, meaning it has no \textit{inductive bias}, or specialization to the top tagging task, that naturally accounts for the variable number of jet constituents.
The DNN uses all 7 of the pre-processed constituent-level quantities described in Section~\ref{sec:jets} as inputs.

\subsection{Energy flow network}
\label{sec:efn}

The Energy Flow Network~\cite{energyflow_networks} (EFN) is a model specifically engineered for jet tagging.
It uses the DeepSets structure~\cite{deepsets}, which ensures permutation invariance across the jet constituents used as input and naturally handles the variable number of jet constituents.
The EFN also uses a \pt weighting mechanism to ensure that lower \pt consituents have lower impact on the output of the network.
This \pt weighting can be interpreted as enforcing infrared and collinear (IRC) safety~\cite{energyflowpoly}.
The output for any of the other networks considered in this study is not IRC safe.
The EFN uses the logarithm of the constituent \pt as input to the \pt weighting mechanism, and does not use the constituent energy as input.

\subsection{Particle flow network}
\label{sec:pfn}

The particle flow network~\cite{energyflow_networks} (PFN) has a very similar structure to the EFN that naturally deals with the variable number of jet constituents and enforces permutation invariance.
However it does not use the \pt weighting mechanism, and uses the constituent energy and all other constituent-level quantities described in Section~\ref{sec:jets} as inputs.

\subsection{ResNet 50}
\label{sec:resnet}

ResNet 50~\cite{resnet} is a large-scale convolutional neural network (CNN) designed for image classification tasks.
CNNs operate on two dimensional arrays whose values give pixel intensity.
Noting the similarity between energy deposits in the ATLAS calorimeter and standard two-dimensional images~\cite{cnn1, cnn2, cnn3, cnn4, cnn5}, the jets are converted into ``jet images'' by binning each constituent's $\eta$ and $\phi$ coordinates into 64 bins, equally spaced in the range of $[-2, 2]$.
The image is then a 64x64 square array where the pixel values are the sum of the raw \pt of the constituents within the pixel, normalized such that the sum of the pixel intensity over the image is one.
Pixel intensities are then rescaled by $\log(1+100\times \pt)$ to make lower \pt patterns in the jet substructure visible.

\begin{figure}[t]
\centering
\subfloat[]{
\includegraphics[width=0.45\textwidth]{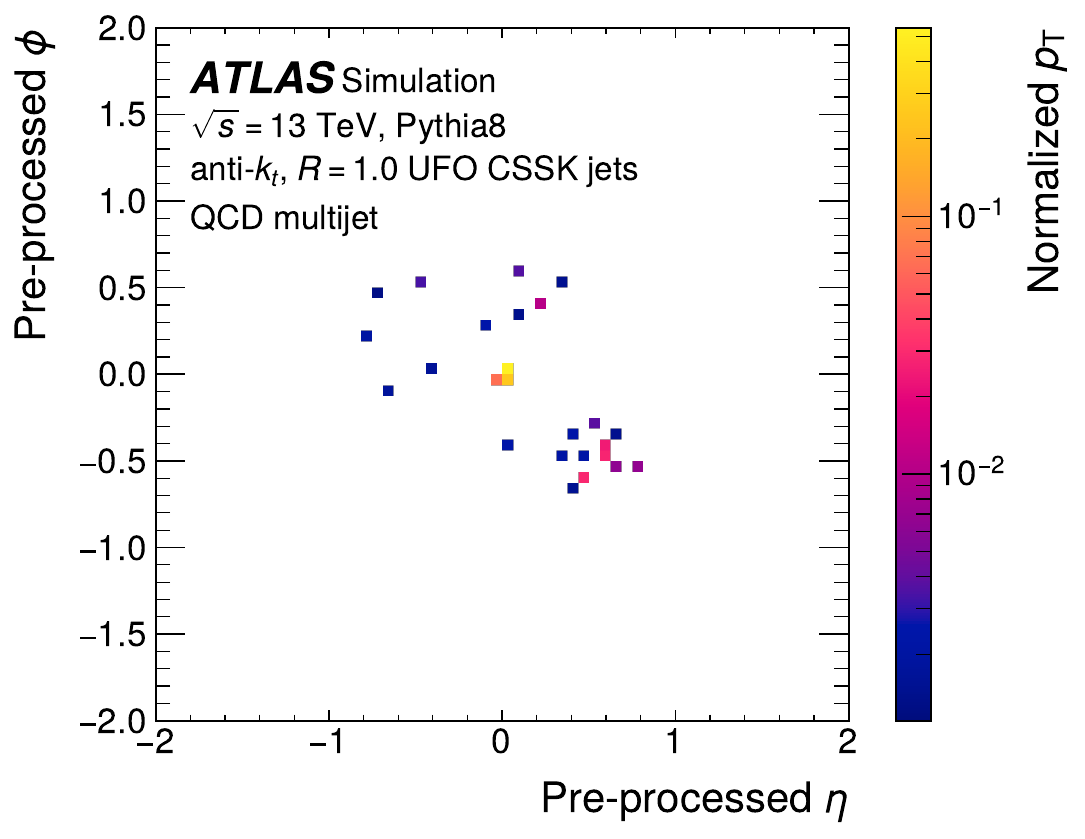}
\label{fig:bkg_jet}
}
\subfloat[]{
\includegraphics[width=0.45\textwidth]{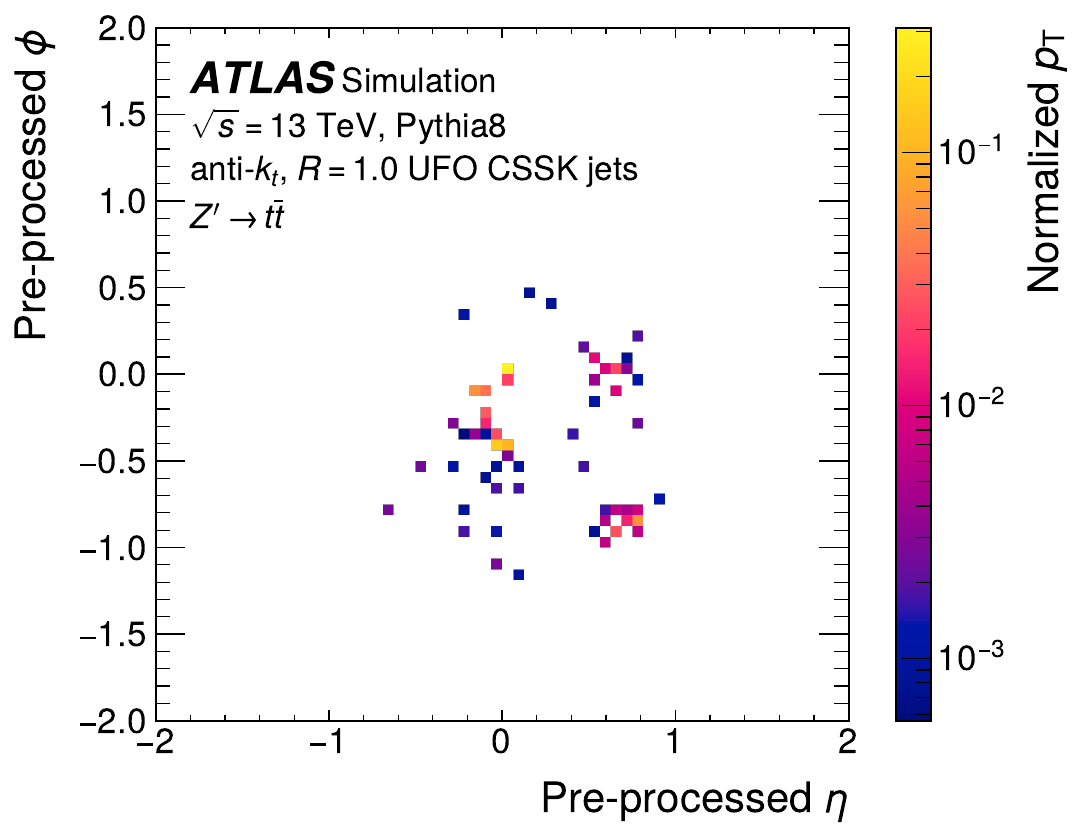}
\label{fig:sig_jet}
} \\
\subfloat[]{
\includegraphics[width=0.45\textwidth]{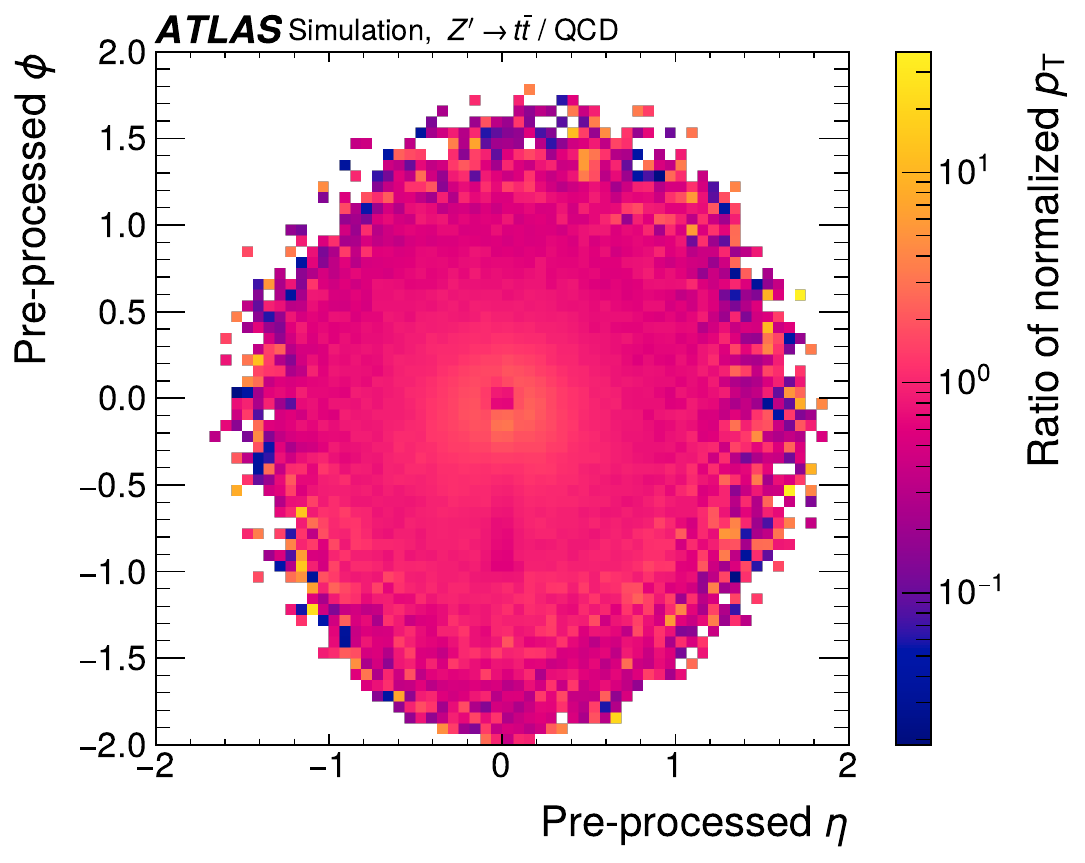}
\label{fig:ratio_jet}
}
\caption{
\protect \subref{fig:bkg_jet} An example background jet image. \protect \subref{fig:sig_jet} An example signal jet image. \protect \subref{fig:ratio_jet} The ratio of the average signal and background jet images.
}
\label{fig:jet_image}
\end{figure}

Typically the images produced for single jets are very sparse, with most pixels containing no constituents.
Example signal and background jet images are shown in Figure \ref{fig:jet_image}, along with an image which shows the ratio of the average signal and background jets.
The differences between the average radiation patterns for signal and background jets can be seen in deviations of the ratio from one.
An excess of transverse momentum concentrated just below the origin on the negative $\phi$-axis results from the second prong of boosted top jets which is preprocessed to align with this axis.
The diffuse excess of transverse momentum distributed around the center is due to the third prong of the boosted top jets.
Finally the deficit of transverse momentum in the center of the image is due to the more collimated nature of light-quark and gluon jets.

\subsection{ParticleNet}
\label{sec:pnet}

ParticleNet~\cite{particle_net} is a graph neural network (GNN) which represents each jet as a graph composed of nodes and edges.
Each constituent in a jet is associated with a node, where all of the input quantities are taken as features of the node.
In this study, these are the 7 constituent-level quantities defined in Section~\ref{sec:jets}.
Each node is connected by an edge to its $k$ nearest neighbors in the $\eta$-$\phi$ plane, where $k$ is a network hyper-parameter.
ParticleNet applies a specialized form of the EdgeConv operation~\cite{edge_conv} to this graph.
This operation is similar to the two dimensional convolution used in CNNs, but defined on graphs instead of images.

Like the EFN and PFN, ParticleNet naturally handles the variable lengths of jets and enforces permutation invariance.
However the EdgeConv operation acts on the feature vectors of pairs of constituents that are spatially close to each other, rather than each constituent separately.
These paired inputs allow ParticleNet to exploit the local relations between constituents.

\FloatBarrier

\section{Tagger performance}
\label{sec:performance}

Performance metrics for the six taggers evaluated on the testing set are shown in Table \ref{tab:performance_numbers}.
The metrics are the area under the receiving-operator-characteristic curve (AUC)~\cite{auc}, the fraction of correct predictions (ACC), and the inverse of the background efficiency (background rejection) at two different working points that fix the signal efficiencies to 50\% and 80\%.
The uncertainties in the performance metrics are the quadrature sum of the uncertainty from the finite size of the testing set (statistical uncertainty), and the uncertainty from the random initialization of the weights and the stochastic nature of network training (training uncertainty).
The statistical uncertainty is calculated as the standard error of the performance metrics over 100 bootstrap replicas of the testing set~\cite{ATL-PHYS-PUB-2021-011}.
The training uncertainty is calculated by training each network 10 times on the same training set with different weight initializations and batching of training data, and then evaluating the standard error of the performance metrics over the 10 training runs.
The differences in the performance metrics between the taggers are larger than the uncertainties in the performance metrics.
In all metrics, ParticleNet achieves the best performance, followed by the PFN, the DNN, and the high-level-quantity-based tagger.
The EFN and ResNet50 fail to outperform the hlDNN, despite access to the constituent information.
The structure of the EFN ensures it is insensitive to low \pt jet constituents, but at the expense of not fully exploiting the available information.
The weaker relative performance of ResNet50 is more suprising, given its strong performance on datasets generated with a parametric detector simulation~\cite{top_tagging_landscape}.
The parametric detector simulation assumes uniform calorimeter cell granularity.
Realistic calorimeters like those used in ATLAS have a non-uniform granularity, which is captured in the high quality simulation used to produce ATLAS simulated data.
Building jet images by applying a uniform pixelization to jet constituents in the context of a non-uniform calorimeter granularity could produce non-physical distortions that are not present in simulated events generated with a parametric detector simulation.
This could explain the weaker performance of ResNet50 on ATLAS simulated data.

The background rejection at the 50\% and 80\% signal efficiency working points is shown as a function of the jet \pt in Figure \ref{fig:background_vs_pt} for the hlDNN, PFN, and ParticleNet taggers.
Unlike the high-level-quantity-based tagger, the constituent-based taggers' performances are best in the mid-\pt range  around 1-2 \TeV.
The high \pt decrease in background rejection is expected, since the higher collimation of jets at high-\pt makes it harder for the tracking detector and calorimeters to resolve the 3-pronged substructure of boosted top jets.
The low-\pt increase in background rejection appears for all constituent-based taggers.

\renewcommand{\arraystretch}{1.5}
\begin{table}[]
\centering
\caption{The performance of each top quark tagger is measured with several metrics evaluated on the testing set. AUC is the area under the receiving-operator-characteristic curve, ACC is the accuracy, and $\varepsilon^{-1}_{bkg}$ is the inverse background efficiency (or background rejection) evaluated at working points which yield a given signal efficiency ($\varepsilon_{sig}$) across the entire testing set. For all metrics, a higher value means better performance, and the table is sorted by increasing AUC. The uncertainty reported on the metrics is the quadrature sum of the uncertainty from the finite statistics of the testing set and the error from the random initialization of network weights and the stochastic nature of network training.}
\begin{tabular}{c|c|c|c|c}
\hline \hline
Tagger & AUC & ACC & $\varepsilon^{-1}_{bkg}$ @ $\varepsilon_{sig} = 0.5$ & $\varepsilon^{-1}_{bkg}$ @ $\varepsilon_{sig} = 0.8$ \\ \hline
ResNet 50 & $0.872 \pm 0.006$ & $0.787 \pm 0.006$ & $18.4 \pm 1.1$ & $4.63 \pm 0.2$ \\
EFN & $0.894 \pm 0.001$ & $0.810 \pm 0.001$ & $23.8 \pm 0.5$ & $5.74 \pm 0.07$ \\
hlDNN & $0.9374 \pm 0.0001$ & $0.8628 \pm 0.0002$ & $47.2 \pm 0.4$ & $10.36 \pm 0.03$ \\
DNN & $0.9447 \pm 0.0004$ & $0.8715 \pm 0.0008$ & $73.0 \pm 1.3$ & $12.5 \pm 0.1$ \\
PFN & $0.9502 \pm 0.0004$ & $0.878 \pm 0.001$ & $92.7 \pm 1.8$ & $14.6 \pm 0.2$ \\
ParticleNet & $0.9614 \pm 0.0005$ & $0.895 \pm 0.001$ & $155.8 \pm 3.8$ & $20.6 \pm 0.4$ \\ \hline \hline
\end{tabular}
\label{tab:performance_numbers}
\end{table}

\begin{figure}
\centering
\subfloat[]{
\includegraphics[width=0.45\textwidth]{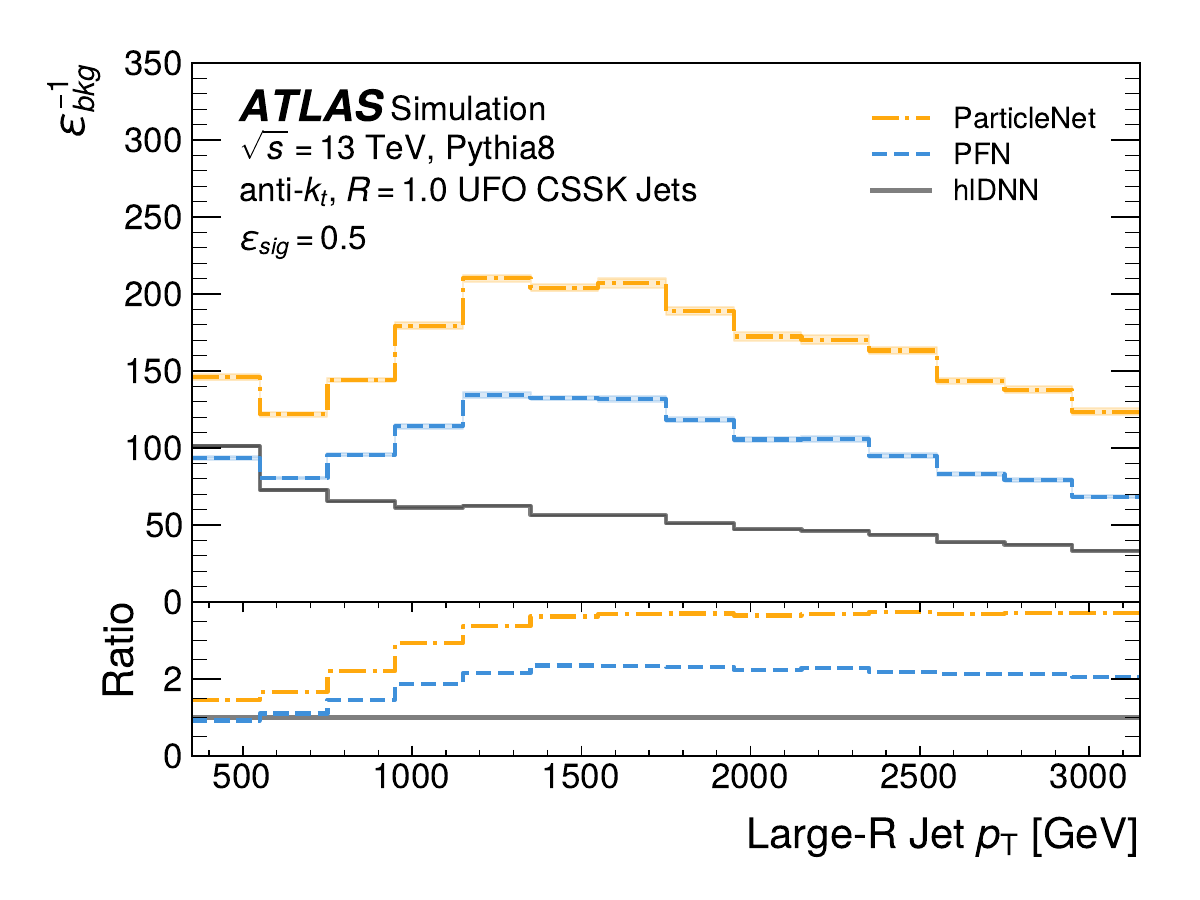}
\label{fig:nominal_bp_50}
}
\subfloat[]{
\includegraphics[width=0.45\textwidth]{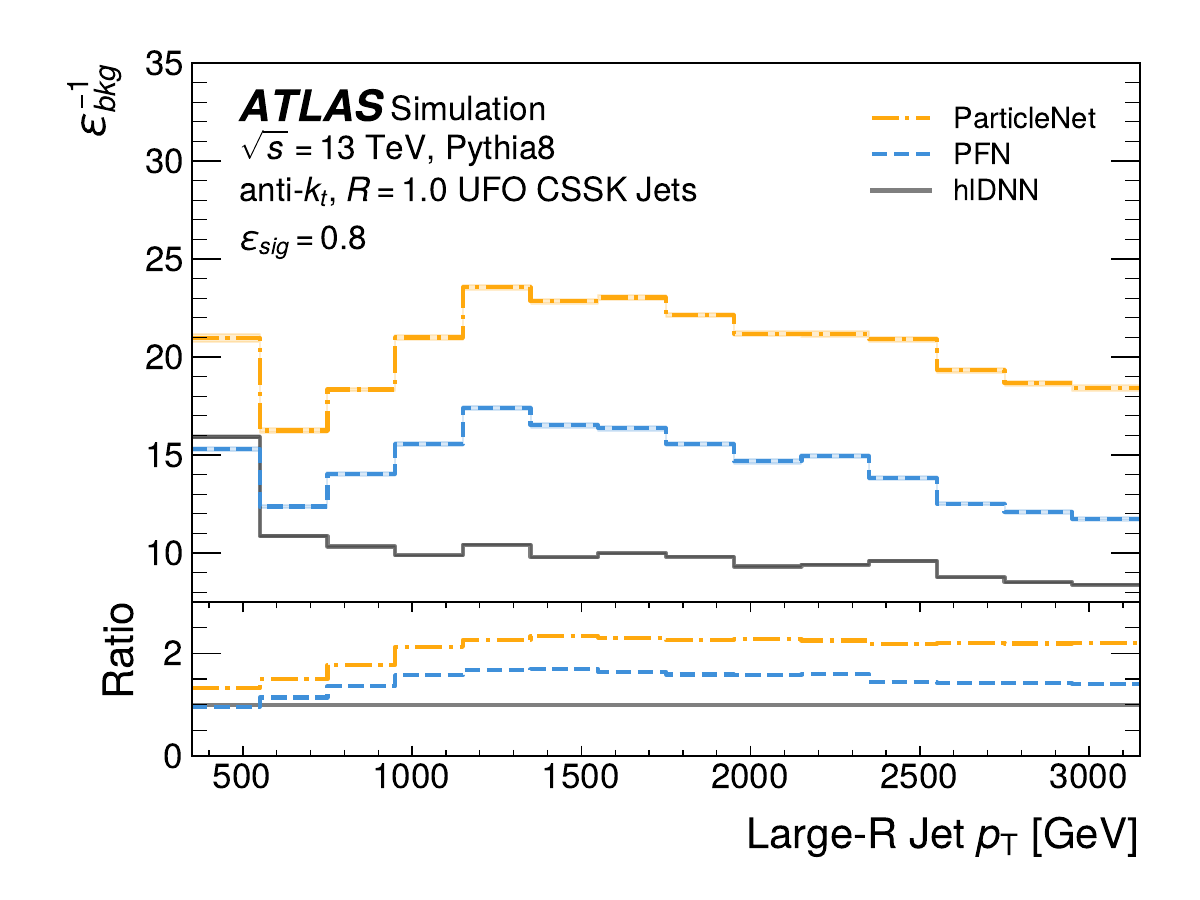}
\label{fig:nominal_bp_80}
}
\caption{Background rejection, or inverse background efficiency ($\varepsilon^{-1}_{bkg}$), of the hlDNN, PFN, and ParticleNet top quark taggers as a function of the jet \pt at the \protect \subref{fig:nominal_bp_50} 50\% signal efficiency and \protect \subref{fig:nominal_bp_80} 80\% signal efficiency working points.  Shaded error bands are the quadrature sum of the error from the finite size of the testing set and the error from the random initialization of network weights and the stochastic nature of network training.}
\label{fig:background_vs_pt}
\end{figure}

\FloatBarrier

\section{Systematic uncertainties}
\label{sec:uncertainties}

Samples of simulated events generated with Monte Carlo methods are a useful model of the experimental data collected by the ATLAS detector for a given physics process, but as important differences between simulated and experimental data can exist, the efficiency of a tagger in experimental data cannot be assumed to be the same as the efficiency in simulated data.
To establish the sensitivity of a physics analysis to a SM or BSM process, this difference in efficiency must be known.
The standard method to establish this difference is to measure a \textit{scale-factor} and the accompaying uncertainties.
Scale-factor measurements are made using samples of signal and background jets collected from experimental data.
These samples are not obtainable for difficult-to-isolate SM or any BSM signature in jet substructure, meaning scale-factors cannot be derived in many applications of jet tagging.
Scale-factor measurements are also time intensive, making it difficult to measure scale factors for two taggers and then compare the size of the scale factors and their uncertainties.

An alternative approach to constraining the difference in tagger efficiency between simulated and experimental data is to apply a set of \textit{systematic variations} directly to simulated data, which alter the tagger inputs within their uncertainties.
These variations produce many systematic varied datasets, and a tagger is then trained on the nominal and evaluated on the nominal and systematic varied datasets.
The differences in tagger efficiency between the nominal and systematic varied datasets can then be used to set an uncertainty in the tagger's efficiency in simulated data.
Provided the systematic variations account for all possible differences between simulated and experimental data at the level of the tagger inputs, the uncertainty provides an estimate of the expected size of the difference in tagger efficiency.

Differences can be classified into two types: possible mis-modeling of the measurements of the kinematic properties of jet constituents by the ATLAS detector (which produce \textit{experimental uncertainties}), and possible mis-modeling of the underlying physical processes that produce the jets (which produce \textit{theoretical uncertainties}).
In this paper, the theoretical uncertainties are assessed by evaluating the tagger efficiency over samples of jets generated with alternative models of the underlying physics processes that produce light quarks, gluons, and top quarks.
This is similar to the procedure used in evaluating theoretical uncertainties in scale-factor measurements~\cite{JETM-2018-03}.
The experimental uncertainties are assessed by varying the kinematic properties of the jet constituents within the uncertainties of the ATLAS detector's measurements.
These uncertainties are established through auxiliary measurements such as Refs.~\cite{PERF-2015-05, JETM-2018-08}.
This is termed the ``bottom-up'' approach to experimental uncertainties, which has been used to produce several measurements of jet substructure observables~\cite{STDM-2017-04, STDM-2017-33, STDM-2017-34}.
In practice it is difficult to construct a set of variations that cover all possible experimental uncertainties without over-covering some uncertainties and disregarding others.
For this reason the approach of applying systematic variations directly to simulated data is generally less precise than scale-factor measurements, but it offers several advantages.
Once the systematically varied datasets are constructed, it is very easy to set uncertainties on the efficiency of an arbitrary tagger, as it only requires running inference over additional datasets.
Further, the approach requires no samples of signal and background jets taken from experimental data.
This is particularly useful for analyses which use a jet tagger to identify BSM physics signatures.

In this section, the standard approach to theoretical uncertainties is combined with the bottom-up approach to set uncertainties on the background rejection of the taggers.
In a realistic physics analysis, scale factors and their uncertainties would be derived for the signal efficiency.
This study derives uncertainties on the background rejection to have a single performance metric and its associated uncertainty which can be compared between taggers.
Several assumptions are made to simplify the experimental uncertainties assigned in this study which will be mentioned explicitly in the following sections.
As a result the experimental uncertainties are only intended to be an estimate of the relative size of the experimental uncertainties associated with each tagger, and are not meant to be used in a physics analysis.

\subsection{Experimental uncertainties}
\label{sec:experimental_uncertainties}

The UFOs used as inputs to the constituent-based taggers can be classified into three types: charged, neutral, and merged~\cite{JETM-2018-06}.
Charged UFOs are simply charged PFOs where an inner detector track~\cite{ATL-PHYS-PUB-2015-051} is matched to a topological cluster~\cite{PERF-2014-07} and used to subtract the expected calorimeter energy from the cluster in a process called \textit{cell subtraction}.
The properties of charged UFOs are determined by the underlying inner detector track, so a set of systematic variations covering track uncertainties are applied to these objects~\cite{STDM-2018-57, STDM-2024-01}.

Neutral UFOs are the topological clusters which remain after both the cell subtraction procedure in the Particle Flow algorithm, and the splitting procedure in the TCC algorithm.
Their properties are primarily determined by the underlying topological cluster, but the inner detector tracks can also affect them through the cell subtraction and TCC splitting procedures.
In this paper, the simplifying assumption is made that neutral UFOs are only affected by the underlying topological cluster, and so a set of systematic variations covering topological cluster uncertainties are applied to these objects~\cite{STDM-2017-33, STDM-2017-34, JETM-2018-08}.

Merged UFOs start as charged PFOs where the cell subtraction procedure is disabled due to a large amount of calorimeter activity in the vicinity of the track.
The TCC algorithm is then run with these charged PFOs as input.
The UFOs this algorithm outputs have their \pt determined by the properties of the underlying topological cluster, and their $\eta$ and $\phi$ determined by the properties of the underlying inner detector track.
Since merged UFOs have their properties set by a combination of the properties of the underlying inner detector tracks and topological clusters, a selection of the track and topological cluster systematic variations are applied to these objects.
This paper makes the simplifying assumption that the $\eta$ and $\phi$ coordinates of merged UFOs are determined solely by the properties of the underlying inner detector track, and the \pt and energy of merged UFOs are determined solely by the properties of the underlying topological cluster.
This is not strictly true because of the complex interplay between the inner detector tracks and topological clusters in the particle flow and TCC algorithms.

The track uncertainties are covered by three systematic variations: the track fake rate, the tracking efficiency, and the track bias.
The track fake rate systematic variation accounts for uncertainty in the rate of tracks produced by chance alignment of signals in the tracking detector.
The size of this uncertainty is estimated by studying the non-linear component of the evolution of the number of inner detector tracks with increasing pile-up in experimental data collected through random triggers~\cite{ATL-PHYS-PUB-2015-051}.
The systematic variation selectively drops charged or merged UFOs, which has the effect of decreasing the track fake rate by its uncertainty.
An increase in the track fake rate is then covered by symmetrizing the uncertainty in the tagger background rejection.

The tracking efficiency systematic variation accounts for uncertainty in the efficiency of finding true tracks.
It contains components that account for limited knowledge of the inner detector material~\cite{ATL-PHYS-PUB-2015-051}, and the merging of tracks within dense tracking environments such as the cores of jets~\cite{PERF-2015-08}.
Like the track fake rate systematic variation, the track efficiency systematic variation drops charged or merged UFOs, but with different probabilities.
This has the effect of decreasing the tracking efficiency, and an increase in the efficiency is covered by symmetrizing the uncertainty in the tagger background rejection.

The track bias systematic variation accounts for possible biases in track \pt measurements due to residual misalignments in the ATLAS tracking detector~\cite{IDTR-2019-05}.
This systematic variation biases the ratio of the track's charge to its momentum for charged UFOs, which in turn shifts the UFO \pt.
It is only applied to charged UFOs, since merged UFOs have their \pt set by the properties of the underlying topological cluster.

The relative uncertainty in background rejection due to the track uncertainties as a function of the jet \pt is shown in Figure \ref{fig:track_uncertainties} for the EFN and ParticleNet taggers.
Fluctuations in the track uncertainties result from the finite statistics of the nominal and systematic varied datasets.
For both taggers the track fake rate is the dominant uncertainty in most \pt bins.
ParticleNet is found to be more sensitive to the track systematic uncertainties than the EFN.

\begin{figure}
\centering
\subfloat[]{
\includegraphics[width=0.45\textwidth]{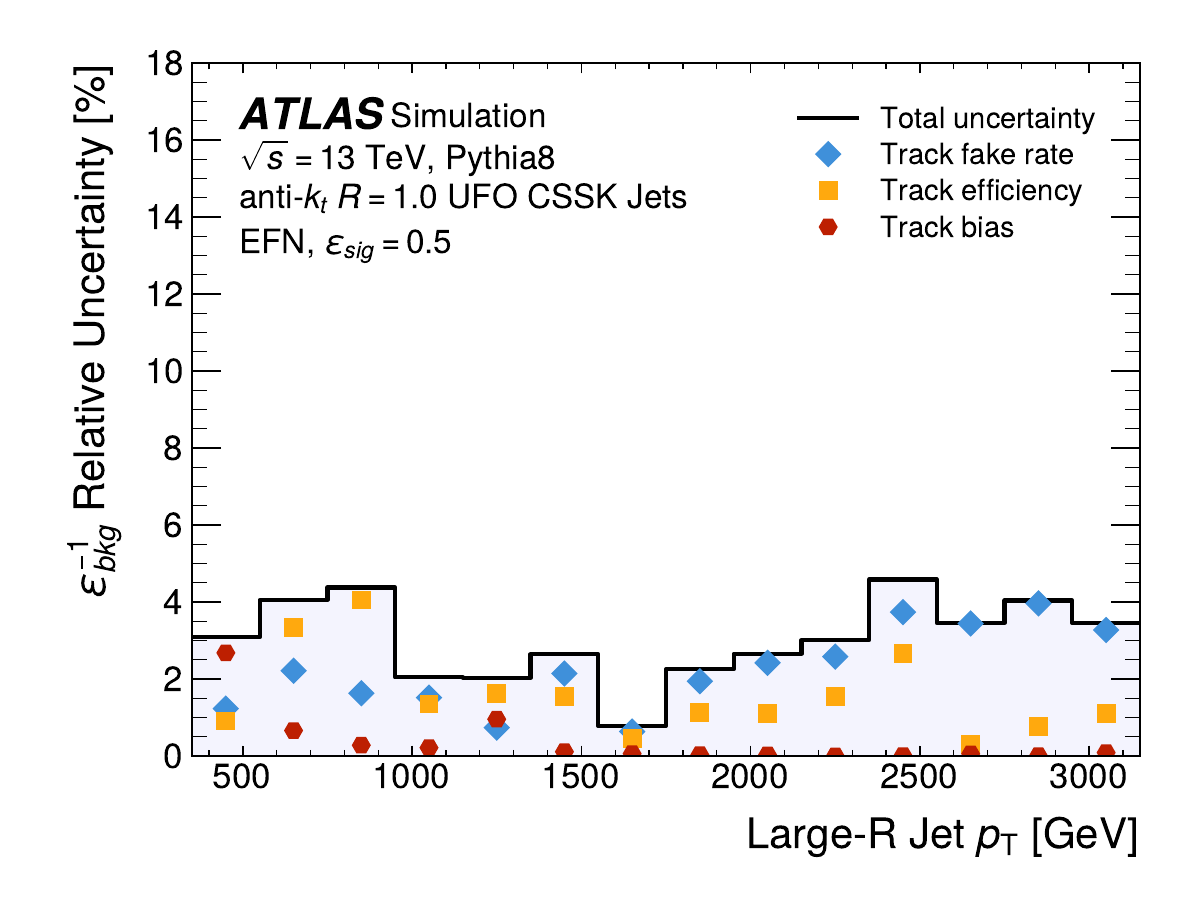}
\label{fig:efn_track_summary}
}
\subfloat[]{
\includegraphics[width=0.45\textwidth]{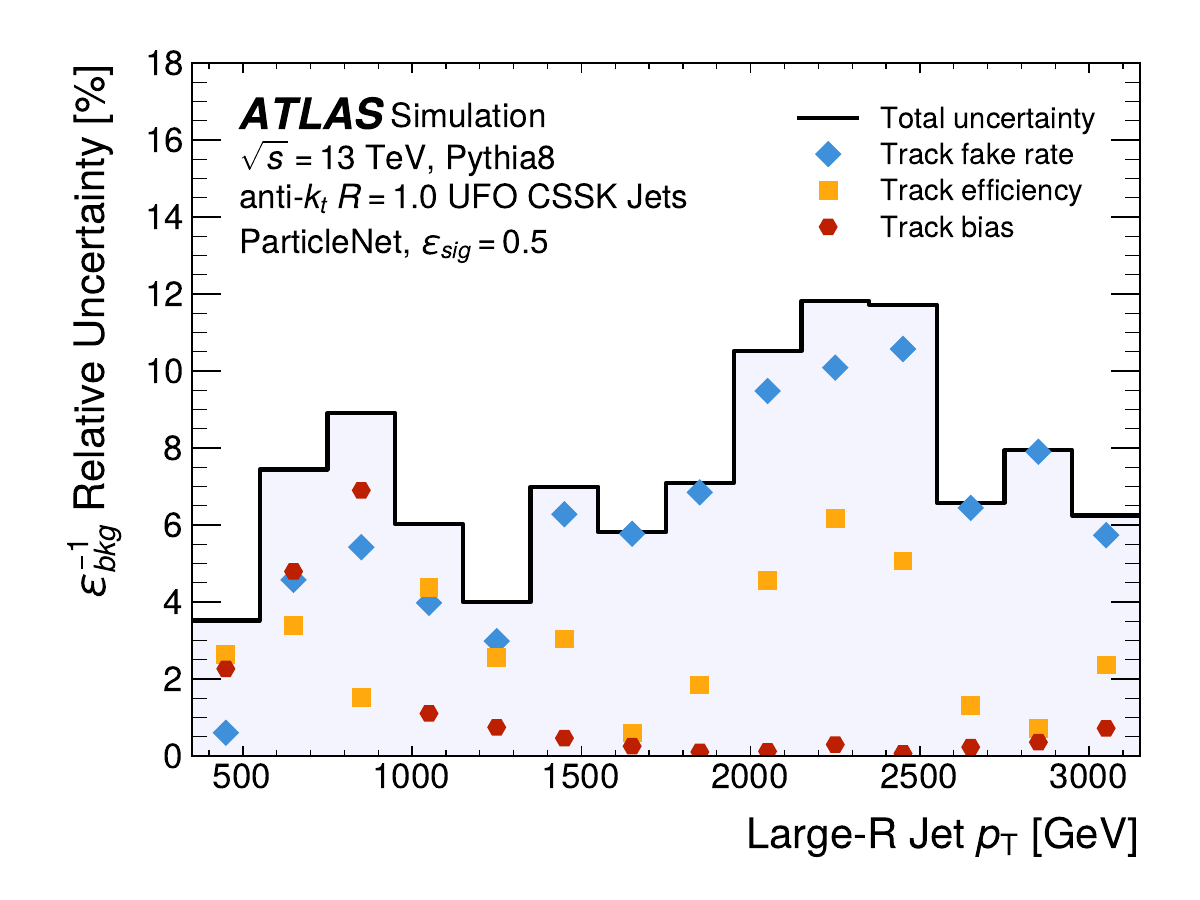}
\label{fig:pnet_track_summary}
}
\caption{The relative uncertainty in the background rejection due to the track systematic uncertainties as a function of the jet \pt for \protect \subref{fig:efn_track_summary} the EFN and \protect \subref{fig:pnet_track_summary} ParticleNet taggers. The total uncertainty is the quadrature sum of all track uncertainties. The fluctuations in the uncertainty result from the finite statistics of the nominal and systematic varied datasets. All uncertainties are evaluated at the 50\% signal efficiency working point.}
\label{fig:track_uncertainties}
\end{figure}

\begin{figure}
\centering
\subfloat[]{
\includegraphics[width=0.45\textwidth]{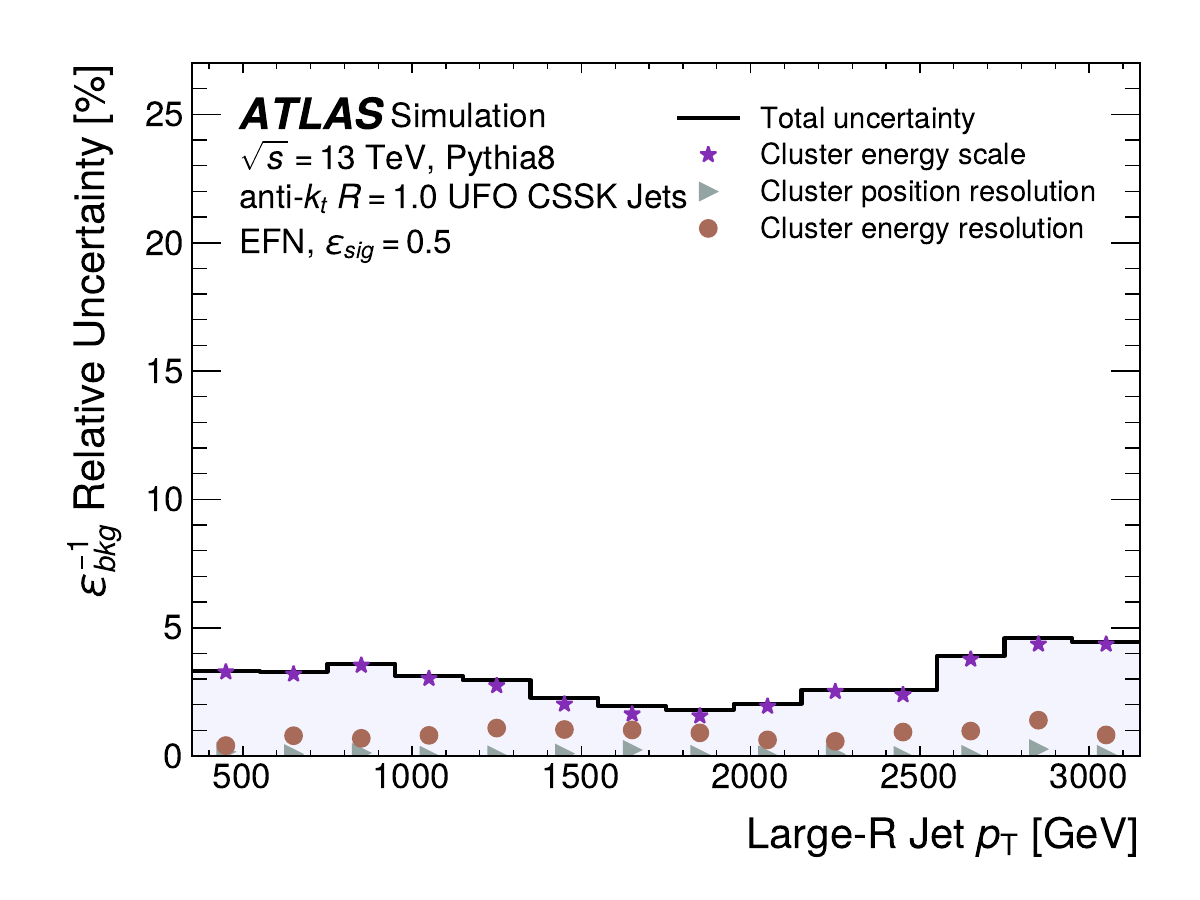}
\label{fig:efn_cluster_summary}
}
\subfloat[]{
\includegraphics[width=0.45\textwidth]{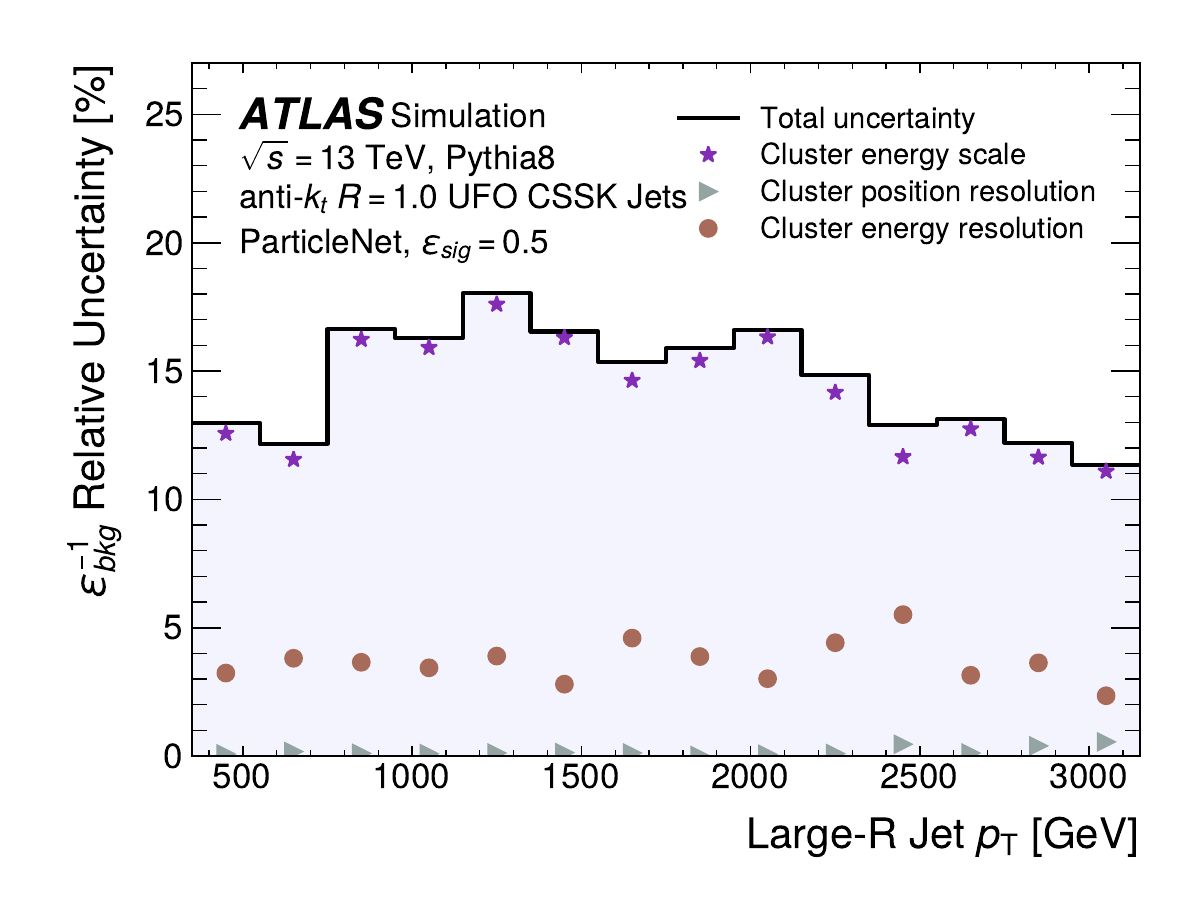}
\label{fig:pnet_cluster_summary}
}
\caption{The relative uncertainty in the background rejection due to the cluster systematic uncertainties as a function of the jet \pt for \protect \subref{fig:efn_cluster_summary} the EFN and \protect \subref{fig:pnet_cluster_summary} ParticleNet taggers. The total uncertainty is the quadrature sum of all cluster uncertainties. All uncertainties are evaluated at the 50\% signal efficiency working point.}
\label{fig:cluster_uncertainties}
\end{figure}

The cluster uncertainties are covered by three systematic variations: the cluster energy scale, the cluster energy resolution, and the cluster position resolution.
The cluster reconstruction efficiency is a negligible source of uncertainty at the \pt scale of the clusters contained within the jets in this study.
The cluster energy scale systematic variation accounts for uncertainty in the response of the ATLAS calorimeter.
A difference in the cluster energy scale between simulated and experimental data would produce a coherent shift in the energy and \pt measurements of topological clusters toward higher or lower energy.
The size of these possible differences is estimated by matching topological clusters to isolated inner detector tracks, and studying the differences between the ratio of the topological cluster's energy to the inner detector track's momentum (this quantity is often termed E/p) between simulated and experimental data~\cite{PERF-2015-05}.
The difference in the mean of the E/p distributions between simulated and experimental data gives the cluster energy scale uncertainty.
The energy and \pt measurements of the neutral and merged UFOs are then shifted either up or down by an amount within the cluster energy scale uncertainty to create the systematic varied datasets.
The final uncertainty is then the maximum of the differences in background rejection between the nominal and the cluster energy scale varied up and down datasets.

The cluster energy resolution systematic variation accounts for uncertainty in the energy resolution of the ATLAS calorimeter.
The size of this uncertainty is estimated by comparing the standard deviations of the distributions of E/p between simulated and experimental data.
The cluster energy resolution is varied up by an amount within the uncertainty by applying Gaussian smearing to the energy and \pt measurements.
It is then varied down by symmetrizing the uncertainty in the tagger efficiency.

Finally, the cluster position resolution systematic variation accounts for uncertainty in the position resolution of the ATLAS calorimeter.
The size of this uncertainty is set by comparing the angular coordinates between isolated tracks matched to isolated clusters.
Like the cluster energy resolution systematic, the cluster position resolution is varied up by applying Gaussian noise, but it acts on the $\eta$ and $\phi$ coordinates instead of the energy and \pt.
It is varied down by symmetrizing the uncertainty in the tagger efficiency.
This variation is applied only to neutral UFOs since the $\eta$ and $\phi$ coordinates of merged UFOs are set by the underlying inner detector track.

In addition to these systematic variations, the taggers may be sensitive to the splitting and merging of topological clusters, where a single particle is reconstructed as two or more clusters, or two or more particles are reconstructed as single cluster.
This paper makes the simplifying assumption that this effect, and any other effects that may bias the reconstruction of jet constituents, is negligible.

The relative uncertainty in background rejection due to the cluster uncertainties as a function of the jet \pt is shown in Figure \ref{fig:cluster_uncertainties} for the EFN and ParticleNet taggers.
For both taggers the cluster energy scale uncertainty is dominant in all \pt bins.

\subsection{Theoretical uncertainties}
\label{sec:theoretical_uncertainties}

Quarks and gluons produced in proton--proton collisions undergo a parton shower and hadronization before they form jets.
Given the strongly coupled nature of QCD, the description of the parton shower and hadronization processes within the MC simulation is not exact.
The possible differences between simulated and experimental data that result are covered by parton shower and hadronization modeling uncertainties.
For the $Z' \rightarrow \ttbar$ process used to obtain a sample of boosted top jets, these uncertainties are estimated by evaluating the tagger background rejection with jets obtained from the two samples of simulated SM \ttbar events described in Section~\ref{sec:mc_simulation}.
These are pure signal jets, so to set an uncertainty in the background rejection the requirements which produce a 50\% signal efficiency are re-calculated using the jets obtained from the SM \ttbar events, and then the background rejection is re-calculated using these requirements.
The difference between the background rejections calculated with the two sets of requirements is taken as the uncertainty.
Since the SM \ttbar process only produces jets in a limited kinematic range, the uncertainty is extrapolated to high \pt by assigning the maximum measured uncertainty to all \pt bins above 1.5 \TeV.

For the QCD multijet process used to obtain light-quark and gluon jets, the uncertainty due to the parton shower and hadronization modeling is estimated by comparing background rejections between the four alternative samples described in Section~\ref{sec:mc_simulation}.
The parton shower modeling uncertainty is estimated by comparing background rejections between the two \HERWIG generated samples which differ only by the parton shower model.
The hadronization modeling uncertainty is likewise estimated by comparing background rejections between the \SHERPA generated samples which differ only by the hadronization model.
In both cases the difference in background rejection between the two samples is taken as the uncertainty.
In a realistic physics analysis involving top quarks, backgrounds would typically be estimated with data-driven methods instead of with simulated data, so the parton shower and hadronization modeling uncertainties for the QCD multijet process would not be relevant.
However the taggers are trained on simulated data, so it is useful to know how the tagger performance is affected by these uncertainties.

The relative uncertainty in background rejection due to the parton shower and hadronization uncertainties as a function of the jet \pt is shown in Figure \ref{fig:pshad_uncertainties} for the EFN and ParticleNet taggers.
Overall the signal process modeling uncertainties are larger than the background process modeling uncertainties.
However the signal process uncertainty covers both the parton shower and hadronization modeling uncertainties together.
The modeling uncertainties for the background process tend to decrease with increasing jet \pt.
The fluctuations in the uncertainties at high \pt, where the uncertainties are small, are due to the finite statistics of the nominal and systematic varied datasets.
It is difficult to discern a trend for the modeling uncertainties for the signal process due to the limited kinematic range of the SM \ttbar process.
As with the experimental uncertainties, ParticleNet is much more sensitive to the parton shower and hadronization modeling uncertainties than the EFN.

\begin{figure}
\centering
\subfloat[]{
\includegraphics[width=0.45\textwidth]{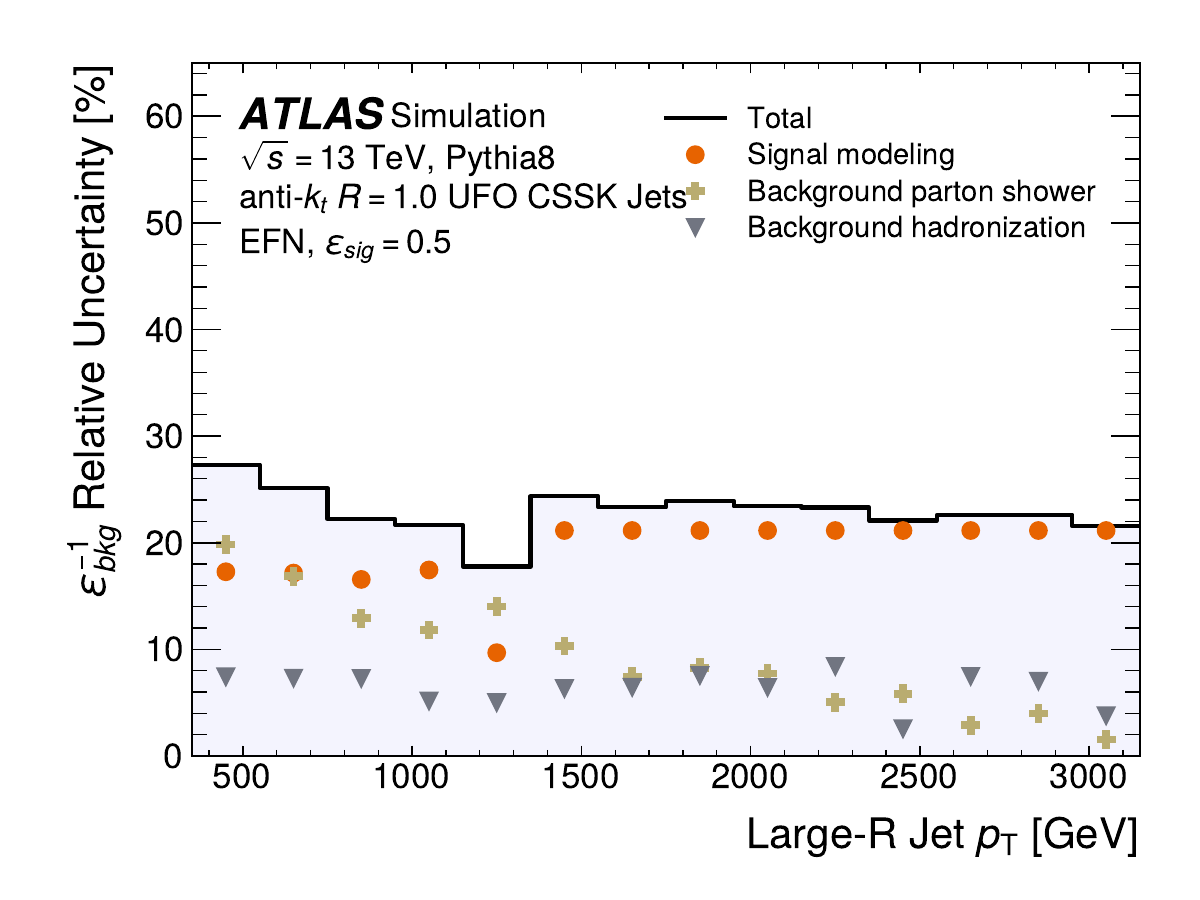}
\label{fig:efn_model_summary}
}
\subfloat[]{
\includegraphics[width=0.45\textwidth]{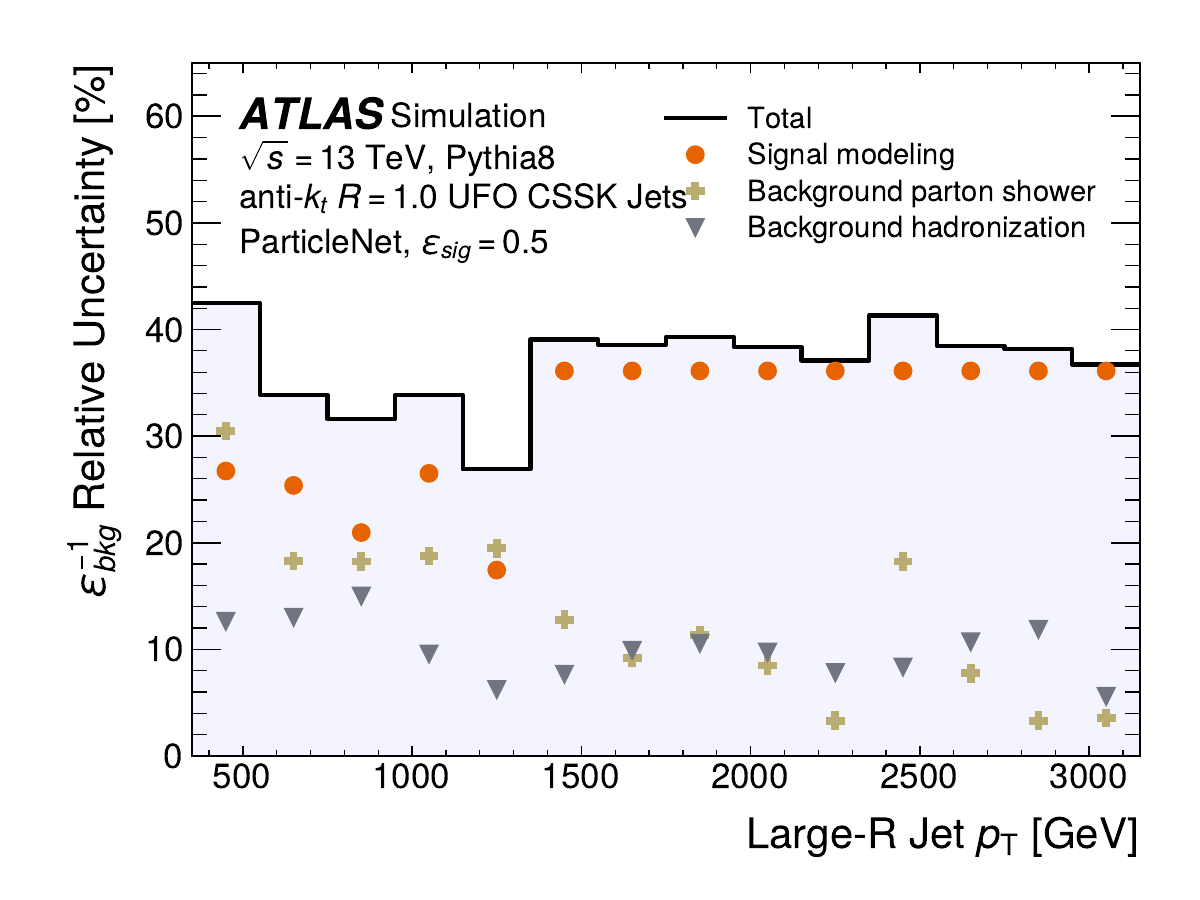}
\label{fig:pnet_model_summary}
}
\caption{The relative uncertainty in the background rejection due to the parton shower and hadronization modeling as a function of the jet \pt for \protect \subref{fig:efn_model_summary} the EFN and \protect \subref{fig:pnet_model_summary} ParticleNet taggers. The total uncertaintiy is the quadrature sum of all parton shower and hadronization modeling uncertainties. All uncertainties are evaluated at the 50\% signal efficiency working point. The uncertainty due to the modeling in the signal process is larger than for the background process, but the signal process uncertainty covers both the parton shower and hadronization modeling uncertainties together.}
\label{fig:pshad_uncertainties}
\end{figure}

\begin{figure}
\centering
\subfloat[]{
\includegraphics[width=0.45\textwidth]{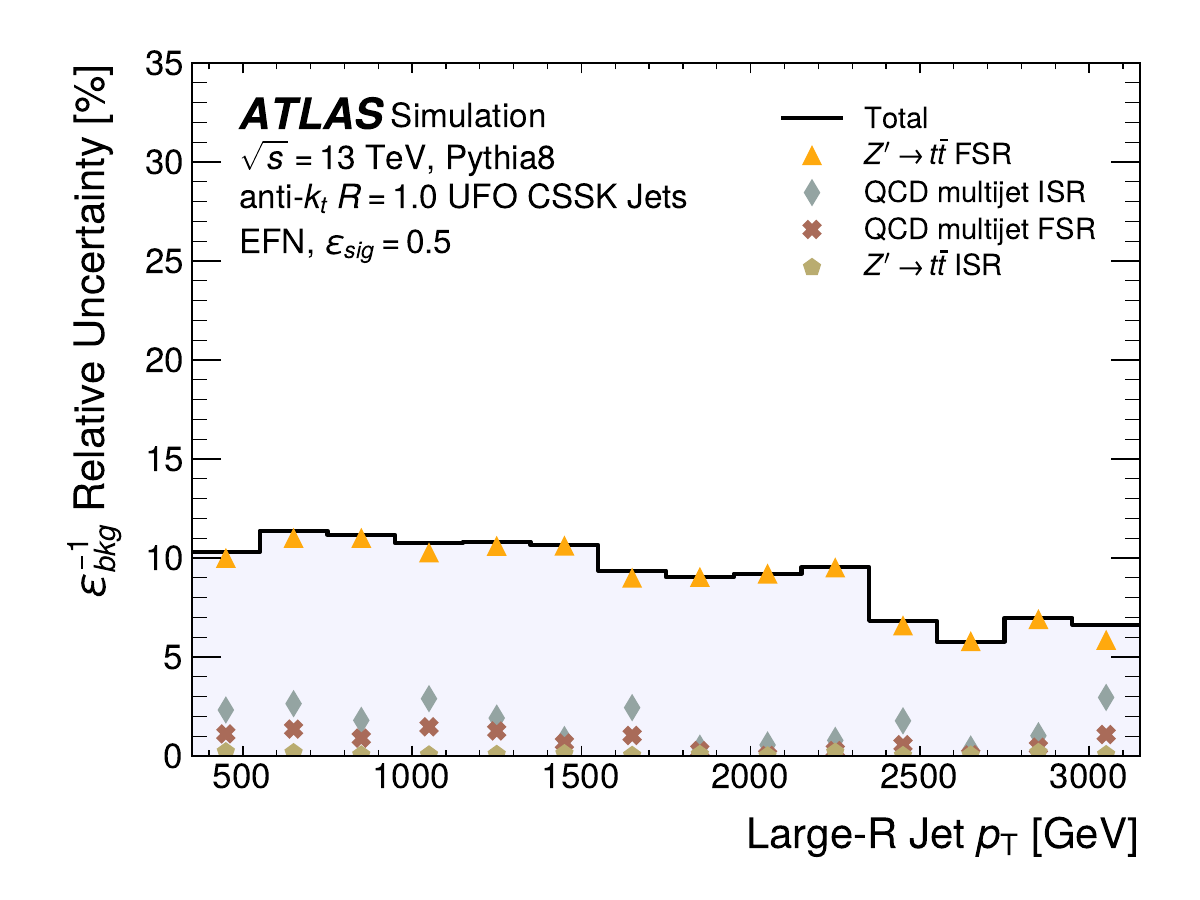}
\label{fig:efn_scales_summary}
}
\subfloat[]{
\includegraphics[width=0.45\textwidth]{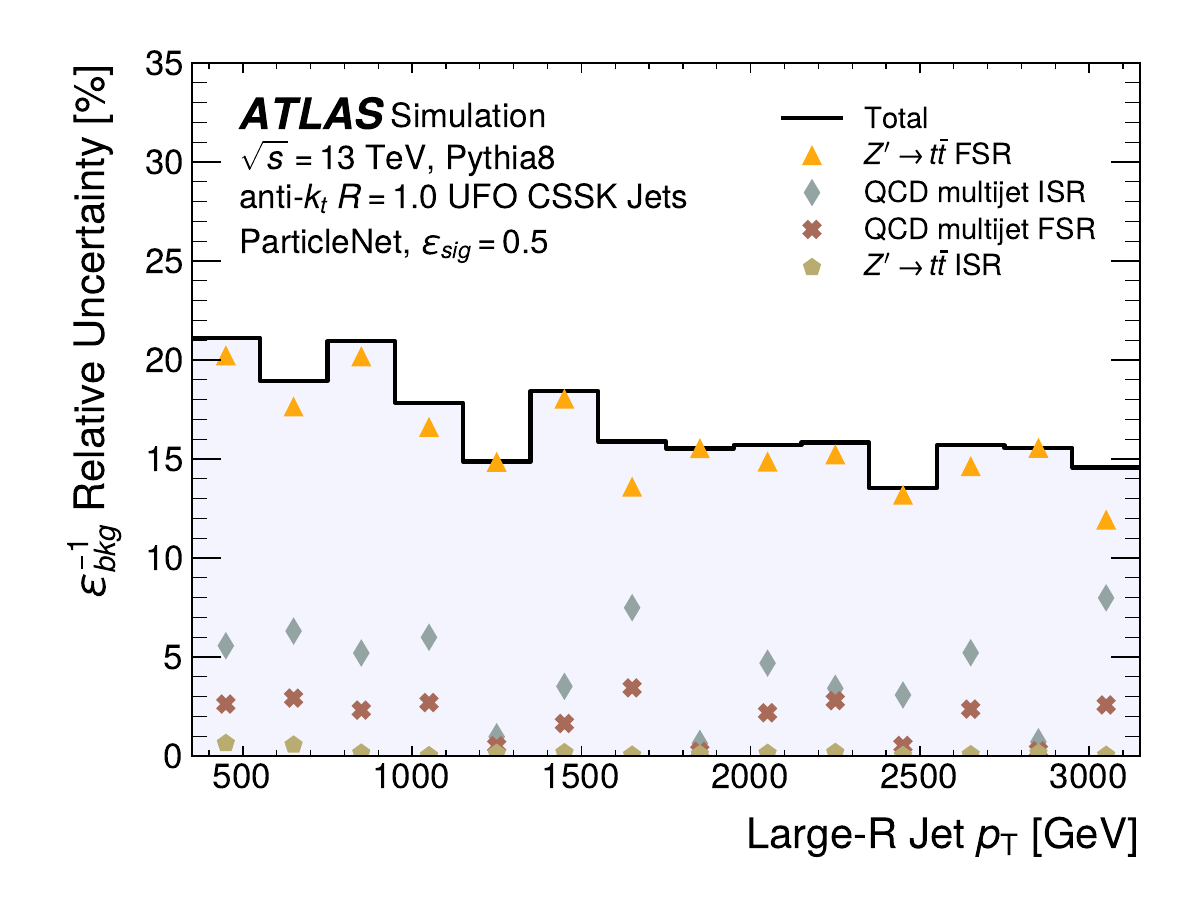}
\label{fig:pnet_scales_summary}
}
\caption{The relative uncertainty in the background rejection due to the choice of renormalization and factorization scales as a function of the jet \pt for \protect \subref{fig:efn_scales_summary} the EFN and \protect \subref{fig:pnet_scales_summary} ParticleNet taggers. The total uncertaintiy is the quadrature sum of all renormalization and factorization scale uncertainties. All uncertainties are evaluated at the 50\% signal efficiency working point.}
\label{fig:renorm_uncertainties}
\end{figure}

The fixed-order matrix element calculations used in the simulation of the $Z' \rightarrow \ttbar$ and QCD multijet processes require a choice of factorization and renormalization scales to remove divergences.
The choice of this scale is unphysical and should not affect the final result of the calculation, but in practice this can happen due to the truncation of the perturbative series.
The scale uncertainties cover possible mis-modeling of the underlying physical processes due to the choice of these scales.
It is assessed by applying \PYTHIA parton shower weights~\cite{pythia_weights} which vary the renormalization and factorization scales for both the initial state radation (ISR) and final state radiation (FSR) up and down by factors of two.
The uncertainty is then estimated by comparing the background rejection of the taggers calculated with the nominal sample to the background rejection calculated with each jet weighted by the relevant shower weight.
The uncertainties are then symmetrized by taking the envelope over the up and down variations.
This process is performed separately for the $Z' \rightarrow \ttbar$ and QCD multijet simulation samples.
The relative uncertainty in the background rejection due to the choice of renormalization and factorization scales as a function of the jet \pt is shown in Figure \ref{fig:renorm_uncertainties} for the EFN and ParticleNet taggers.
The fluctuations in these uncertainties are due to the finite statistics of the testing set.
Again, ParticleNet is associated with larger uncertainties than the EFN.
Both taggers show much larger sensitivity to the FSR scales used in generating the $Z' \rightarrow \ttbar$ sample compared to the other scales.

\subsection{Validation of uncertainties}
\label{sec:validation}

\begin{figure}
\centering
\subfloat[]{
\includegraphics[width=0.45\textwidth]{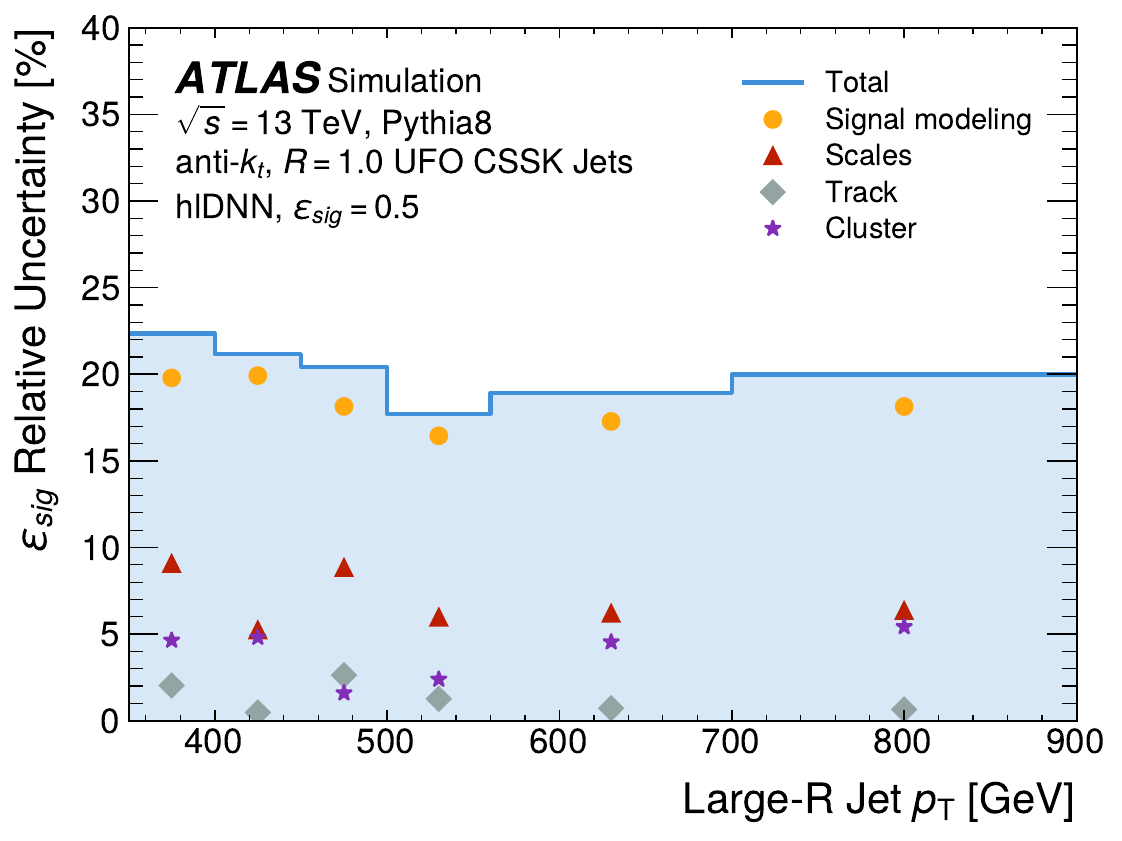}
\label{fig:hlsf_summary}
}
\subfloat[]{
\includegraphics[width=0.45\textwidth]{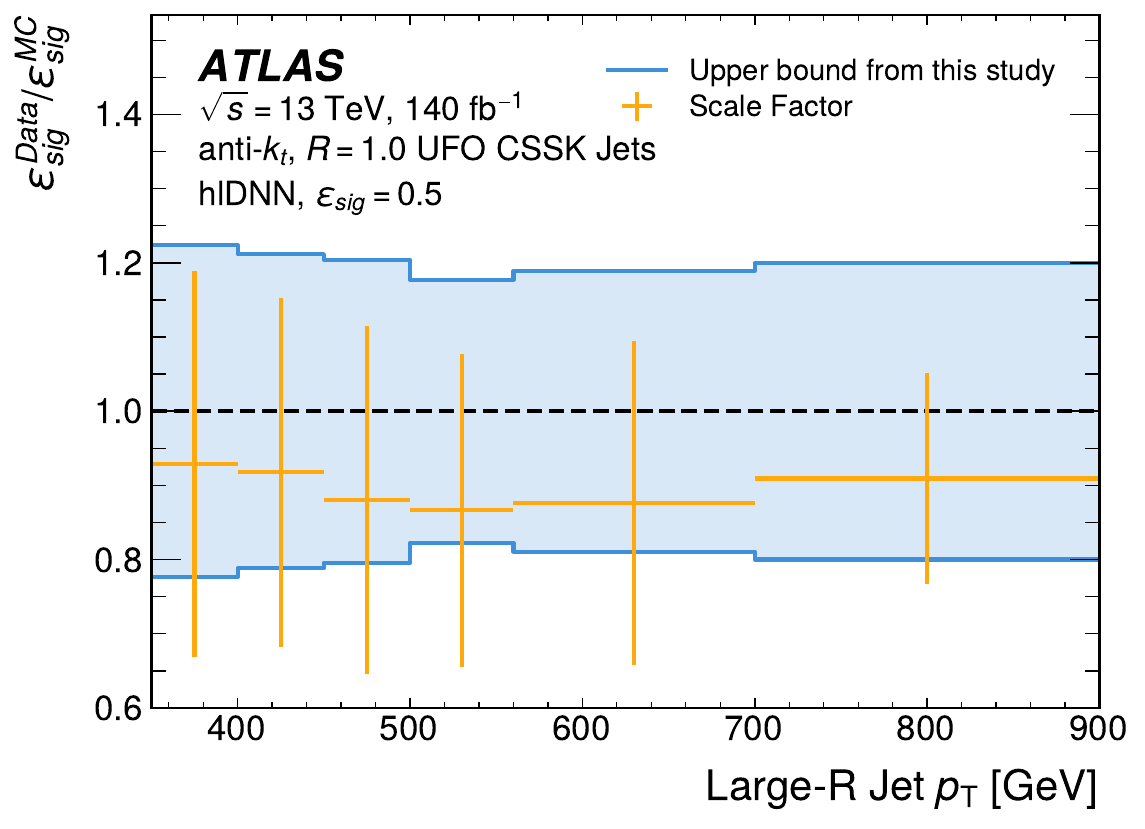}
\label{fig:hlsf_compare}
}
\caption{\protect \subref{fig:hlsf_summary} The uncertainty in the signal efficiency for the hlDNN tagger, calculated as the quadrature sum of all uncertainties considered in this study except the parton shower and hadronization modeling uncertainty for the QCD multijet process. \protect \subref{fig:hlsf_compare} A comparison between the total uncertainty derived in this study and the scale factor and its uncertainty for the hlDNN tagger. The scale factor is measured using the same methods as described in Ref.~\cite{JETM-2018-03}}
\label{fig:validation}
\end{figure}

To place the uncertainties derived in this study in the context of the scale factor approach utilized by realistic physics analyses, a comparison is made between the two methods for the hlDNN tagger.
For comparing uncertainties between taggers, it is useful to calculate uncertainties on the background rejection so that a single metric can be used to characterize both tagger performance and the size of the uncertainties.
However physics analyses targeting boosted top-quarks are interested in the uncertainties on the signal efficiency, since backgrounds are typically estimated with data-driven methods.
Therefore boosted top tagger scale factors are derived for the signal efficiency, so the uncertainties in this study must be transferred to this quantity for a direct comparison.
This excludes the parton shower and hadronization modeling uncertainty for the QCD multijet process, since it is not relevant for the signal efficiency.
The uncertainty budget for the signal efficiency is shown for the hlDNN tagger in Figure \ref{fig:validation}\subref{fig:hlsf_summary}.
Within the jet \pt range of $[350-1000]$ GeV in which scale factors are derived, the parton shower and hadronization modeling uncertainty for the $Z' \rightarrow \ttbar$ process is the dominant uncertainty.

To calculate the scale factor, the signal efficiency in experimental data is measured using the highest \pt large-radius jet in events taken from a semi-leptonic \ttbar enriched signal region.
The uncertainty in the scale factor is calculated by propagating experimental and theoretical uncertainties through the template fit used to extract the signal efficiency.
For details on scale-factor measurements see Ref.~\cite{JETM-2018-03}.
The scale factor is a measurement of the ratio of the signal efficiencies in simulated and experimental data, along with the uncertainty in this ratio.
In contrast the uncertainties derived in this study provide an upper bound, or maximum possible value, for the magnitude of the difference in tagger efficiency between simulated and experimental data.
The deviation of the scale factors from one should then be smaller than the uncertainties dervied in this study.
Figure \ref{fig:validation}\subref{fig:hlsf_compare} shows a comparison between the upper bound on the magnitude of the difference in efficiency derived in this study and the scale factor and its uncertainty for the hlDNN tagger.
The upper bound covers the central value of the scale factor in all \pt bins, providing good validation of the approach.

\subsection{Total uncertainties}
\label{sec:total_uncertainties}

The total uncertainty, calculated as the quadrature sum of all sources of uncertainty described above, is shown as a function of the jet \pt for the EFN and ParticleNet taggers in Figure \ref{fig:total_budgets}.
For both the EFN and ParticleNet, the parton shower and hadronization modeling uncertainties for the $Z' \rightarrow \ttbar$ process are the largest in most \pt bins.
These are followed by the factorization and renormalization scales and background parton shower and hadronization modeling uncertainties.
The total uncertainty as a function of the jet \pt is shown for all constituent-based taggers in Figure \ref{fig:global_errors}\subref{fig:budget_compare}.
All taggers have the largest uncertainties at low jet \pt where the theoretical uncertainties produce large and equal contributions to the total uncertainty.
Figure \ref{fig:global_errors}\subref{fig:error_vs_perf} shows the total uncertainty in the background rejection across the entire \pt range plotted against the background rejection for all taggers considered.
A clear correlation between background rejection and uncertainty in the background rejection is visible, with the most powerful taggers producing the largest uncertainties.
The exception to this correlation is ResNet50, which is associated with large uncertainties while also having poor performance.

\begin{figure}
\centering
\subfloat[]{
\includegraphics[width=0.45\textwidth]{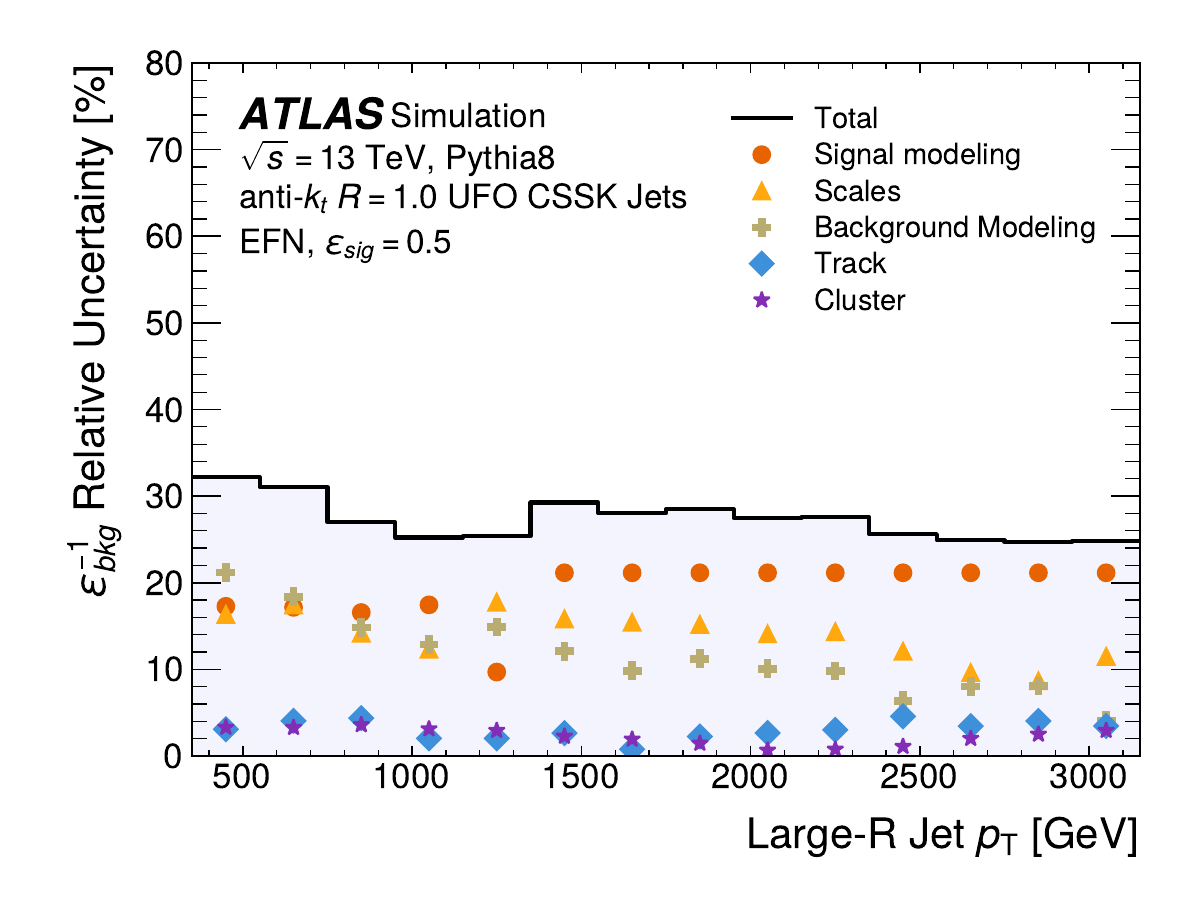}
\label{fig:efn_budget}
}
\subfloat[]{
\includegraphics[width=0.45\textwidth]{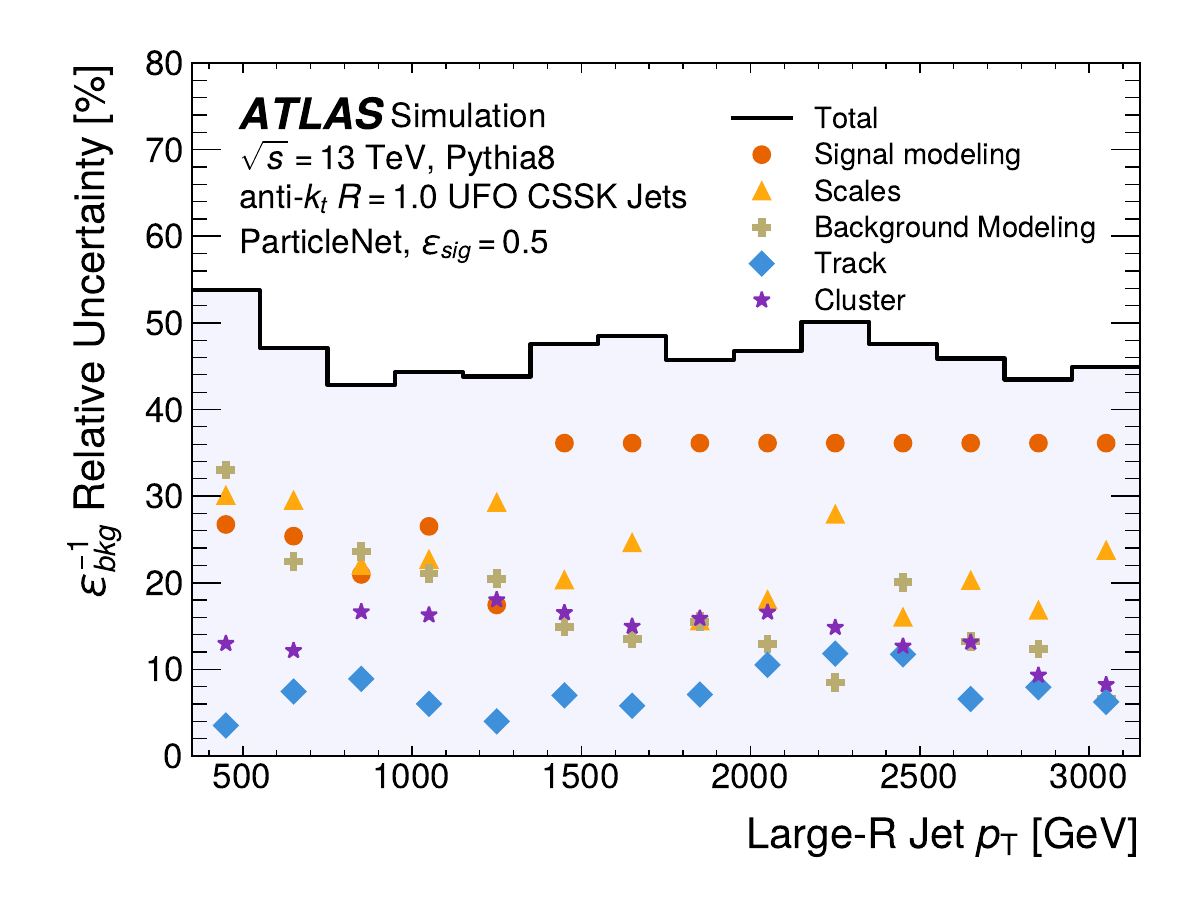}
\label{fig:pnet_budget}
}
\caption{The total uncertainty budget in the background rejection as a function of the jet \pt for \protect \subref{fig:efn_budget} the EFN and \protect \subref{fig:pnet_budget} ParticleNet taggers. All uncertainties are evaluated at the 50\% signal efficiency working point.}
\label{fig:total_budgets}
\end{figure}

\begin{figure}
\centering
\subfloat[]{
\includegraphics[width=0.45\textwidth]{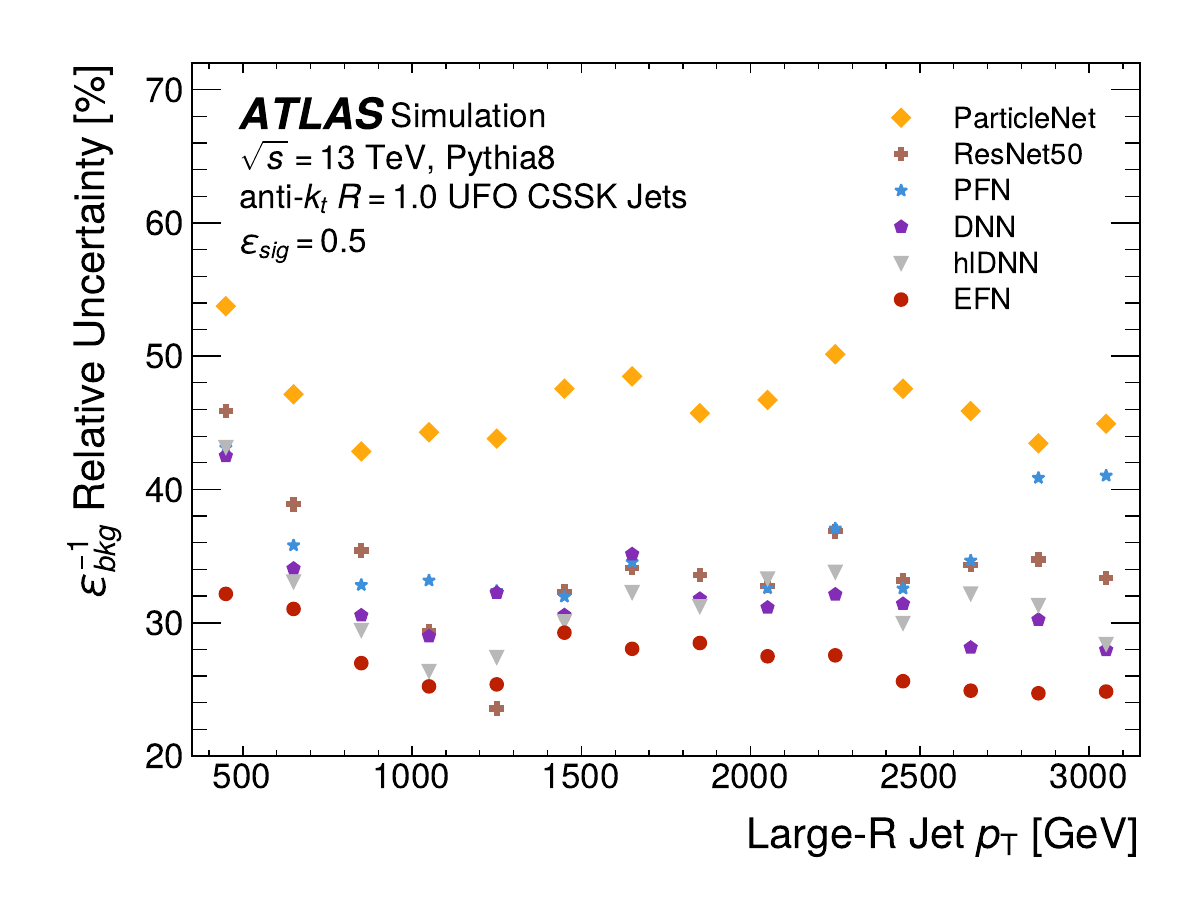}
\label{fig:budget_compare}
}
\subfloat[]{
\includegraphics[width=0.45\textwidth]{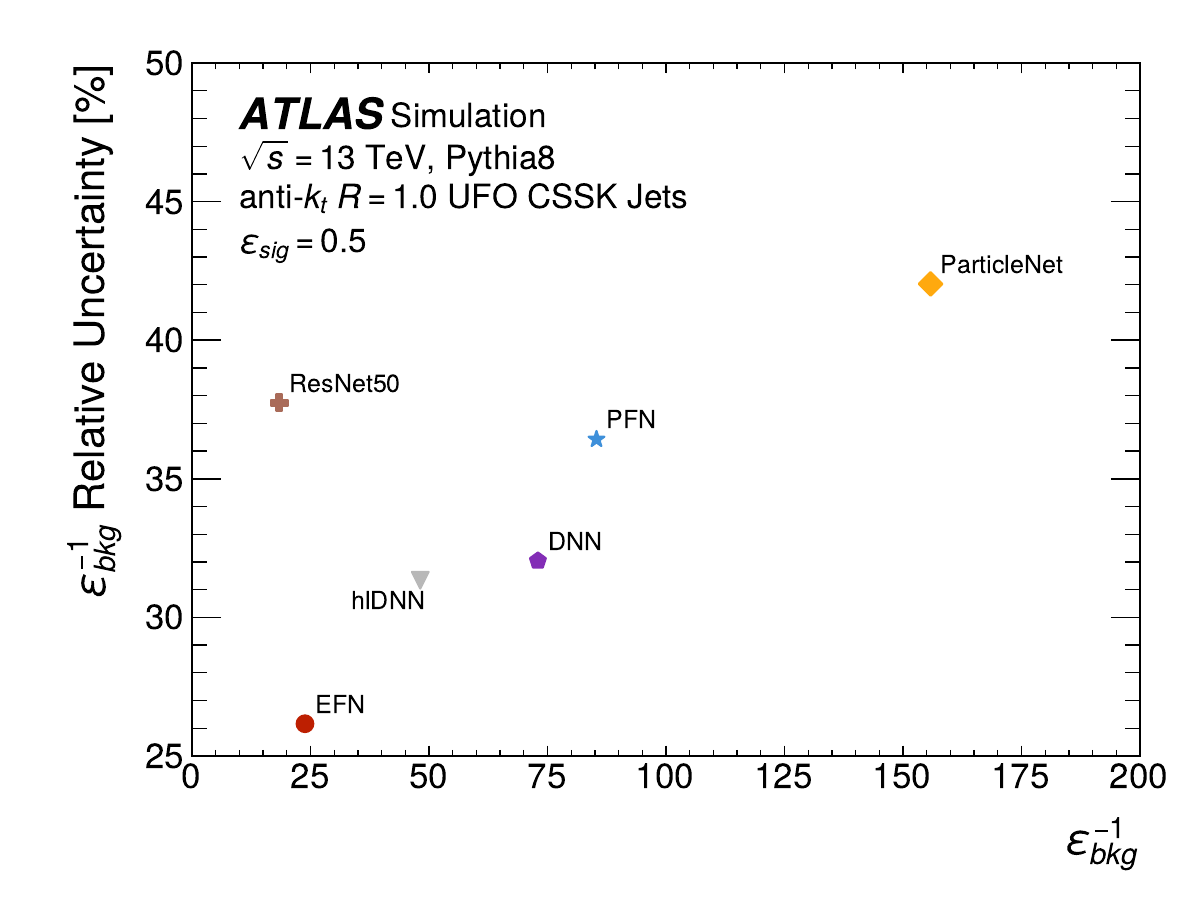}
\label{fig:error_vs_perf}
}
\caption{\protect \subref{fig:budget_compare} The total uncertainty for each of the constituent-based taggers considered in this study, and \protect \subref{fig:error_vs_perf} the total uncertainty in the background rejection across the entire testing set plotted against the background rejection for each constituent-based tagger. All uncertainties are evaluated at the 50\% signal efficiency working point.}
\label{fig:global_errors}
\end{figure}

\FloatBarrier

\section{Conclusion}
\label{sec:conclusion}

This paper presents the performance of selected constituent-based jet taggers on a top tagging task.
Several constituent-based taggers (DNN, PFN, and ParticleNet) outperform the high-level-quantity-based baseline tagger, while the EFN and ResNet50 underperform.
This underperformance was not observed in studies performed with a parametric detector simulation~\cite{top_tagging_landscape}, highlighting the importance of developing taggers in the context of realistic detector simulation.

The systematic uncertainties that would result from the application of taggers to experimental data are then probed by applying systematic variations directly to the simulated data.
A strong correlation between tagger performance and the size of the uncertainties is observed.
The theoretical uncertainties were found to be dominant for all taggers considered.
In particular the parton shower and hadronization modeling uncertainties are dominant in most \pt bins for all taggers.
Compared with the theoretical uncertainties, the experimental uncertainties are found to be small.
The uncertainties derived in this study are not a scale-factor measurement, or uncertainties on a scale factor.
The experimental uncertainties rely on simplifying assumptions which ignore possible differences between simulated and experimental data.
They should be considered as an illustration of the bottom-up approach to experimental systematic uncertainties, rather than a set of uncertainties that could be used in a physics analysis.
However they are useful for establishing the sensitivity of the different taggers to possible differences between simulated and experimental data.

The systematic uncertainties derived in this study stem from performance differences between the nominal and systematic varied datasets.
Therefore large uncertainties suggest that large performance differences between the nominal datasets and experimental data are possible.
The highly performant jet taggers that produce large systematic uncertainties in this study could have lower performance in experimental data than in simulated data, meaning they would not deliver the sensitivity improvements expected from performance studies which do not consider systematic uncertainties.
Even if these performance differences are accounted for by scale-factor measurements, such a degradation in tagger performance is not recoverable.
Additionally scale-factor measurements also have uncertainties which are expected to correlate with the uncertainties derived in this study.
This was verified by comparing the size of the uncertainties derived for the hlDNN in this study to scale factors and scale-factor uncertainties derived using semi-leptonic \ttbar events with the methods described in Ref.~\cite{JETM-2018-03}, and finding good agreement.
Scale-factor uncertainties will be included as a systematic uncertainty in any physics analysis which uses a jet tagger, so larger scale-factor uncertainties will also degrade the sensitivity improvements from a more performant tagger.

The size of scale-factor uncertainties relative to other uncertainties in a physics analysis will be highly dependent on the context, so it is difficult to predict if larger scale-factor uncertainties would be limiting for a given physics analysis.
In some cases larger scale-factor uncertainties could present a limitation in the sensitivity improvements expected from ML-based jet tagging techniques.
Reducing the size of these systematic uncertainties without compromising the tagger performance is then an important direction for future research.
Two possible directions are to design taggers which are robust against, or aware of systematic effects~\cite{isnotallyouneed, pelican1, pelican2, pelican3, lorentznet1, lorentznet2, uncertainty_aware, adversarial_training, perf_vs_resilience, cautionary_tale}, or to train taggers directly on experimental data in a weakly supervised setting~\cite{cwola, therightway, CMS-TOP-18-011}.
To support progress in the first direction and to allow a more realistic assessment of the performance of other existing taggers, the datasets used in this study are made publicly available~\cite{open_data}.
This includes all nominal and systematic varied datasets, including the alternative Monte Carlo samples used to assess the parton shower and hadronization modeling uncertainties.
Additional documentation is provided, which details how to use the datasets to assess the uncertainties associated with an arbitrary tagger.

\section*{Acknowledgements}
%

%
%

%
%

We thank CERN for the very successful operation of the LHC and its injectors, as well as the support staff at
CERN and at our institutions worldwide without whom ATLAS could not be operated efficiently.

The crucial computing support from all WLCG partners is acknowledged gratefully, in particular from CERN, the ATLAS Tier-1 facilities at TRIUMF/SFU (Canada), NDGF (Denmark, Norway, Sweden), CC-IN2P3 (France), KIT/GridKA (Germany), INFN-CNAF (Italy), NL-T1 (Netherlands), PIC (Spain), RAL (UK) and BNL (USA), the Tier-2 facilities worldwide and large non-WLCG resource providers. Major contributors of computing resources are listed in Ref.~\cite{ATL-SOFT-PUB-2023-001}.

We gratefully acknowledge the support of ANPCyT, Argentina; YerPhI, Armenia; ARC, Australia; BMWFW and FWF, Austria; ANAS, Azerbaijan; CNPq and FAPESP, Brazil; NSERC, NRC and CFI, Canada; CERN; ANID, Chile; CAS, MOST and NSFC, China; Minciencias, Colombia; MEYS CR, Czech Republic; DNRF and DNSRC, Denmark; IN2P3-CNRS and CEA-DRF/IRFU, France; SRNSFG, Georgia; BMBF, HGF and MPG, Germany; GSRI, Greece; RGC and Hong Kong SAR, China; ISF and Benoziyo Center, Israel; INFN, Italy; MEXT and JSPS, Japan; CNRST, Morocco; NWO, Netherlands; RCN, Norway; MNiSW, Poland; FCT, Portugal; MNE/IFA, Romania; MESTD, Serbia; MSSR, Slovakia; ARIS and MVZI, Slovenia; DSI/NRF, South Africa; MICIU/AEI, Spain; SRC and Wallenberg Foundation, Sweden; SERI, SNSF and Cantons of Bern and Geneva, Switzerland; NSTC, Taipei; TENMAK, T\"urkiye; STFC/UKRI, United Kingdom; DOE and NSF, United States of America.

Individual groups and members have received support from BCKDF, CANARIE, CRC and DRAC, Canada; CERN-CZ, FORTE and PRIMUS, Czech Republic; COST, ERC, ERDF, Horizon 2020, ICSC-NextGenerationEU and Marie Sk{\l}odowska-Curie Actions, European Union; Investissements d'Avenir Labex, Investissements d'Avenir Idex and ANR, France; DFG and AvH Foundation, Germany; Herakleitos, Thales and Aristeia programmes co-financed by EU-ESF and the Greek NSRF, Greece; BSF-NSF and MINERVA, Israel; NCN and NAWA, Poland; La Caixa Banking Foundation, CERCA Programme Generalitat de Catalunya and PROMETEO and GenT Programmes Generalitat Valenciana, Spain; G\"{o}ran Gustafssons Stiftelse, Sweden; The Royal Society and Leverhulme Trust, United Kingdom.

In addition, individual members wish to acknowledge support from Armenia: Yerevan Physics Institute (FAPERJ); CERN: European Organization for Nuclear Research (CERN PJAS); Chile: Agencia Nacional de Investigaci\'on y Desarrollo (FONDECYT 1230812, FONDECYT 1230987, FONDECYT 1240864); China: Chinese Ministry of Science and Technology (MOST-2023YFA1605700), National Natural Science Foundation of China (NSFC - 12175119, NSFC 12275265, NSFC-12075060); Czech Republic: Czech Science Foundation (GACR - 24-11373S), Ministry of Education Youth and Sports (FORTE CZ.02.01.01/00/22\_008/0004632), PRIMUS Research Programme (PRIMUS/21/SCI/017); EU: H2020 European Research Council (ERC - 101002463); European Union: European Research Council (ERC - 948254, ERC 101089007), Horizon 2020 Framework Programme (MUCCA - CHIST-ERA-19-XAI-00), European Union, Future Artificial Intelligence Research (FAIR-NextGenerationEU PE00000013), Italian Center for High Performance Computing, Big Data and Quantum Computing (ICSC, NextGenerationEU); France: Agence Nationale de la Recherche (ANR-20-CE31-0013, ANR-21-CE31-0013, ANR-21-CE31-0022, ANR-22-EDIR-0002), Investissements d'Avenir Labex (ANR-11-LABX-0012); Germany: Baden-Württemberg Stiftung (BW Stiftung-Postdoc Eliteprogramme), Deutsche Forschungsgemeinschaft (DFG - 469666862, DFG - CR 312/5-2); Italy: Istituto Nazionale di Fisica Nucleare (ICSC, NextGenerationEU), Ministero dell'Università e della Ricerca (PRIN - 20223N7F8K - PNRR M4.C2.1.1); Japan: Japan Society for the Promotion of Science (JSPS KAKENHI JP22H01227, JSPS KAKENHI JP22H04944, JSPS KAKENHI JP22KK0227, JSPS KAKENHI JP23KK0245); Netherlands: Netherlands Organisation for Scientific Research (NWO Veni 2020 - VI.Veni.202.179); Norway: Research Council of Norway (RCN-314472); Poland: Ministry of Science and Higher Education (IDUB AGH, POB8, D4 no 9722), Polish National Agency for Academic Exchange (PPN/PPO/2020/1/00002/U/00001), Polish National Science Centre (NCN 2021/42/E/ST2/00350, NCN OPUS nr 2022/47/B/ST2/03059, NCN UMO-2019/34/E/ST2/00393, UMO-2020/37/B/ST2/01043, UMO-2021/40/C/ST2/00187, UMO-2022/47/O/ST2/00148, UMO-2023/49/B/ST2/04085); Slovenia: Slovenian Research Agency (ARIS grant J1-3010); Spain: Generalitat Valenciana (Artemisa, FEDER, IDIFEDER/2018/048), Ministry of Science and Innovation (MCIN \& NextGenEU PCI2022-135018-2, MICIN \& FEDER PID2021-125273NB, RYC2019-028510-I, RYC2020-030254-I, RYC2021-031273-I, RYC2022-038164-I), PROMETEO and GenT Programmes Generalitat Valenciana (CIDEGENT/2019/027); Sweden: Swedish Research Council (Swedish Research Council 2023-04654, VR 2018-00482, VR 2022-03845, VR 2022-04683, VR 2023-03403, VR grant 2021-03651), Knut and Alice Wallenberg Foundation (KAW 2018.0157, KAW 2018.0458, KAW 2019.0447, KAW 2022.0358); Switzerland: Swiss National Science Foundation (SNSF - PCEFP2\_194658); United Kingdom: Leverhulme Trust (Leverhulme Trust RPG-2020-004), Royal Society (NIF-R1-231091); United States of America: U.S. Department of Energy (ECA DE-AC02-76SF00515), Neubauer Family Foundation.

%
%


%
%
%
%
%
%
%

%
%
%
%
%
%

%
%
%

%

%
%
%
%
%
%
%
%
%
%
%
%
%
%
%

%

%
%
%
\printbibliography

@article{top_tagging_landscape,
    author = "Kasieczka, Gregor and others",
    title = "{The Machine Learning landscape of top taggers}",
    eprint = "1902.09914",
    archivePrefix = "arXiv",
    primaryClass = "hep-ph",
    doi = "10.21468/SciPostPhys.7.1.014",
    journal = "SciPost Phys.",
    volume = "7",
    pages = "014",
    year = "2019"
}

@article{pythia8,
    author = {Sj\"ostrand, Torbj\"orn and Ask, Stefan and Christiansen, Jesper R. and Corke, Richard and Desai, Nishita and Ilten, Philip and Mrenna, Stephen and Prestel, Stefan and Rasmussen, Christine O. and Skands, Peter Z.},
    title = "{An introduction to PYTHIA 8.2}",
    eprint = "1410.3012",
    archivePrefix = "arXiv",
    primaryClass = "hep-ph",
    reportNumber = "LU-TP-14-36, MCNET-14-22, CERN-PH-TH-2014-190, FERMILAB-PUB-14-316-CD, DESY-14-178, SLAC-PUB-16122",
    doi = "10.1016/j.cpc.2015.01.024",
    journal = "Comput. Phys. Commun.",
    volume = "191",
    pages = "159--177",
    year = "2015"
}

@article{constituent_subtraction1,
    author = "Berta, Peter and Spousta, Martin and Miller, David W. and Leitner, Rupert",
    title = "{Particle-level pileup subtraction for jets and jet shapes}",
    eprint = "1403.3108",
    archivePrefix = "arXiv",
    primaryClass = "hep-ex",
    doi = "10.1007/JHEP06(2014)092",
    journal = "JHEP",
    volume = "06",
    pages = "092",
    year = "2014"
}

@article{constituent_subtraction2,
    author = "Berta, P. and Masetti, L. and Miller, D. W. and Spousta, M.",
    title = "{Pileup and Underlying Event Mitigation with Iterative Constituent Subtraction}",
    eprint = "1905.03470",
    archivePrefix = "arXiv",
    primaryClass = "hep-ph",
    doi = "10.1007/JHEP08(2019)175",
    journal = "JHEP",
    volume = "08",
    pages = "175",
    year = "2019"
}

@article{soft_killer,
    author = "Cacciari, Matteo and Salam, Gavin P. and Soyez, Gregory",
    title = "{SoftKiller, a particle-level pileup removal method}",
    eprint = "1407.0408",
    archivePrefix = "arXiv",
    primaryClass = "hep-ph",
    reportNumber = "CERN-PH-TH-2014-116",
    doi = "10.1140/epjc/s10052-015-3267-2",
    journal = "Eur. Phys. J. C",
    volume = "75",
    number = "2",
    pages = "59",
    year = "2015"
}

@article{soft_drop,
    author = "Larkoski, Andrew J. and Marzani, Simone and Soyez, Gregory and Thaler, Jesse",
    title = "{Soft Drop}",
    eprint = "1402.2657",
    archivePrefix = "arXiv",
    primaryClass = "hep-ph",
    reportNumber = "MIT-CTP-4531, DCPT-14-24, IPPP-14-12",
    doi = "10.1007/JHEP05(2014)146",
    journal = "JHEP",
    volume = "05",
    pages = "146",
    year = "2014"
}

@article{anti-kt,
    author = "Cacciari, Matteo and Salam, Gavin P. and Soyez, Gregory",
    title = "{The anti-$k_t$ jet clustering algorithm}",
    eprint = "0802.1189",
    archivePrefix = "arXiv",
    primaryClass = "hep-ph",
    reportNumber = "LPTHE-07-03",
    doi = "10.1088/1126-6708/2008/04/063",
    journal = "JHEP",
    volume = "04",
    pages = "063",
    year = "2008"
}

@article{resnet,
  title={Deep Residual Learning for Image Recognition},
  author={He, Kaiming and Zhang, Xiangyu and Ren, Shaoqing and Sun, Jian},
  booktitle={Proceedings of the IEEE conference on computer vision and pattern recognition},
  year={2015},
  archivePrefix = "arXiv",
  eprint = "1512.03385"
}

@article{energyflow_networks,
    author = "Komiske, Patrick T. and Metodiev, Eric M. and Thaler, Jesse",
    title = "{Energy Flow Networks: Deep Sets for Particle Jets}",
    eprint = "1810.05165",
    archivePrefix = "arXiv",
    primaryClass = "hep-ph",
    reportNumber = "MIT-CTP 5064",
    doi = "10.1007/JHEP01(2019)121",
    journal = "JHEP",
    volume = "01",
    pages = "121",
    year = "2019"
}

@article{particle_net,
    author = "Qu, Huilin and Gouskos, Loukas",
    title = "{Jet Tagging via Particle Clouds}",
    eprint = "1902.08570",
    archivePrefix = "arXiv",
    primaryClass = "hep-ph",
    doi = "10.1103/PhysRevD.101.056019",
    journal = "Phys. Rev. D",
    volume = "101",
    number = "5",
    pages = "056019",
    year = "2020"
}

@article{edge_conv,
  title={Dynamic Graph CNN for Learning on Point Clouds},
  author={Wang, Yue and Sun, Yongbin and Liu, Ziwei and Sarma, Sanjay E and Bronstein, Michael M and Solomon, Justin M},
  journal={ACM Transactions On Graphics},
  volume={38},
  number={5},
  year={2019},
  publisher={ACM New York, NY, USA},
  archivePrefix = "arXiv",
  eprint = "1801.07829",
  doi = "10.1145/3326362"
}

@article{herwig1,
    author = "Bellm, Johannes and others",
    title = "{Herwig 7.0/Herwig++ 3.0 release note}",
    eprint = "1512.01178",
    archivePrefix = "arXiv",
    primaryClass = "hep-ph",
    reportNumber = "CERN-PH-TH-2015-289, MAN-HEP-2015-15, IFJPAN-IV-2015-13, KA-TP-18-2015, DCPT-15-142, MCNET-15-28, IPPP-15-71, HERWIG-2015-01",
    doi = "10.1140/epjc/s10052-016-4018-8",
    journal = "Eur. Phys. J. C",
    volume = "76",
    number = "4",
    pages = "196",
    year = "2016"
}

@article{herwig2,
    author = "B{\"a}hr, M. and others",
    title = "{Herwig++ Physics and Manual}",
    eprint = "0803.0883",
    archivePrefix = "arXiv",
    primaryClass = "hep-ph",
    reportNumber = "CERN-PH-TH-2008-038, CAVENDISH-HEP-08-03, KA-TP-05-2008, DCPT-08-22, IPPP-08-11, CP3-08-05",
    doi = "10.1140/epjc/s10052-008-0798-9",
    journal = "Eur. Phys. J. C",
    volume = "58",
    pages = "639--707",
    year = "2008"
}

@article{cnn2,
    author = "Baldi, Pierre and Bauer, Kevin and Eng, Clara and Sadowski, Peter and Whiteson, Daniel",
    title = "{Jet Substructure Classification in High-Energy Physics with Deep Neural Networks}",
    eprint = "1603.09349",
    archivePrefix = "arXiv",
    primaryClass = "hep-ex",
    doi = "10.1103/PhysRevD.93.094034",
    journal = "Phys. Rev. D",
    volume = "93",
    number = "9",
    pages = "094034",
    year = "2016"
}

@article{cnn1,
    author = "Cogan, Josh and Kagan, Michael and Strauss, Emanuel and Schwarztman, Ariel",
    title = "{Jet-Images: Computer Vision Inspired Techniques for Jet Tagging}",
    eprint = "1407.5675",
    archivePrefix = "arXiv",
    primaryClass = "hep-ph",
    doi = "10.1007/JHEP02(2015)118",
    journal = "JHEP",
    volume = "02",
    pages = "118",
    year = "2015"
}

@article{boost2012,
   title={Boosted objects and jet substructure at the LHC. Report of BOOST2012, held at IFIC Valencia, 23rd–27th of July 2012},
   eprint = "1311.2708",
   archivePrefix = "arXiv",
   volume={74},
   DOI={10.1140/epjc/s10052-014-2792-8},
   number={3},
   journal={Eur. Phys. J. C},
   publisher={Springer Science and Business Media LLC},
   author={Altheimer, A. and Arce, A. and Asquith, L. and Backus Mayes, J. and Kuutmann, E. Bergeaas and Berger, J. and Bjergaard, D. and Bryngemark, L. and Buckley, A. and Butterworth, J. and Cacciari, M. and Campanelli, M. and Carli, T. and Chala, M. and Chapleau, B. and Chen, C. and Chou, J. P. and Cornelissen, Th. and Curtin, D. and Dasgupta, M. and Davison, A. and de Almeida Dias, F. and de Cosa, A. and de Roeck, A. and Debenedetti, C. and Doglioni, C. and Ellis, S. D. and Fassi, F. and Ferrando, J. and Fleischmann, S. and Freytsis, M. and Gonzalez Silva, M. L. and de la Hoz, S. Gonzalez and Guescini, F. and Han, Z. and Hook, A. and Hornig, A. and Izaguirre, E. and Jankowiak, M. and Juknevich, J. and Kaci, M. and Kar, D. and Kasieczka, G. and Kogler, R. and Larkoski, A. and Loch, P. and Lopez Mateos, D. and Marzani, S. and Masetti, L. and Mateu, V. and Miller, D. W. and Mishra, K. and Nef, P. and Nordstrom, K. and Oliver Garcia, E. and Penwell, J. and Pilot, J. and Plehn, T. and Rappoccio, S. and Rizzi, A. and Rodrigo, G. and Safonov, A. and Salam, G. P. and Salt, J. and Schaetzel, S. and Schioppa, M. and Schmidt, A. and Scholtz, J. and Schwartzman, A. and Schwartz, M. D. and Segala, M. and Son, M. and Soyez, G. and Spannowsky, M. and Stewart, I. and Strom, D. and Swiatlowski, M. and Sanchez Martinez, V. and Takeuchi, M. and Thaler, J. and Thompson, E. N. and Tran, N. V. and Vermilion, C. and Villaplana, M. and Vos, M. and Wacker, J. and Walsh, J.},
   year={2014},
   month={Mar},
   pages={2792},
}

@article{ecf_angles,
    author = "Moult, Ian and Necib, Lina and Thaler, Jesse",
    title = "{New Angles on Energy Correlation Functions}",
    eprint = "1609.07483",
    archivePrefix = "arXiv",
    primaryClass = "hep-ph",
    reportNumber = "MIT-CTP-4825",
    doi = "10.1007/JHEP12(2016)153",
    journal = "JHEP",
    volume = "12",
    pages = "153",
    year = "2016"
}

@article{ecf,
    author = "Larkoski, Andrew J. and Salam, Gavin P. and Thaler, Jesse",
    title = "{Energy Correlation Functions for Jet Substructure}",
    eprint = "1305.0007",
    archivePrefix = "arXiv",
    primaryClass = "hep-ph",
    reportNumber = "MIT-CTP-4446, CERN-PH-TH-2013-066, LPN13-026",
    doi = "10.1007/JHEP06(2013)108",
    journal = "JHEP",
    volume = "06",
    pages = "108",
    year = "2013"
}

@article{ecf_power,
    author = "Larkoski, Andrew J. and Moult, Ian and Neill, Duff",
    title = "{Power Counting to Better Jet Observables}",
    eprint = "1409.6298",
    archivePrefix = "arXiv",
    primaryClass = "hep-ph",
    reportNumber = "MIT--CTP-4588",
    doi = "10.1007/JHEP12(2014)009",
    journal = "JHEP",
    volume = "12",
    pages = "009",
    year = "2014"
}

@article{nsubjet,
    author = "Thaler, Jesse and Van Tilburg, Ken",
    title = "{Identifying Boosted Objects with N-subjettiness}",
    eprint = "1011.2268",
    archivePrefix = "arXiv",
    primaryClass = "hep-ph",
    reportNumber = "MIT-CTP-4191",
    doi = "10.1007/JHEP03(2011)015",
    journal = "JHEP",
    volume = "03",
    pages = "015",
    year = "2011"
}

@article{nsubjet2,
    author = "Thaler, Jesse and Van Tilburg, Ken",
    title = "{Maximizing Boosted Top Identification by Minimizing N-subjettiness}",
    eprint = "1108.2701",
    archivePrefix = "arXiv",
    primaryClass = "hep-ph",
    reportNumber = "MIT-CTP-4287",
    doi = "10.1007/JHEP02(2012)093",
    journal = "JHEP",
    volume = "02",
    pages = "093",
    year = "2012"
}

@article{energyflowpoly,
    author = "Komiske, Patrick T. and Metodiev, Eric M. and Thaler, Jesse",
    title = "{Energy flow polynomials: A complete linear basis for jet substructure}",
    eprint = "1712.07124",
    archivePrefix = "arXiv",
    primaryClass = "hep-ph",
    reportNumber = "MIT-CTP-4965",
    doi = "10.1007/JHEP04(2018)013",
    journal = "JHEP",
    volume = "04",
    pages = "013",
    year = "2018"
}

@article{CMS-PAS-BTV-22-001,
      author = "{CMS Collaboration}",
      title         = "{Performance of heavy-flavour jet identification in
                       boosted topologies in proton-proton collisions at $\sqrt{s}
                       = 13~\mathrm{TeV}$}",
      institution   = "CERN",
      reportNumber  = "CMS-PAS-BTV-22-001",
      address       = "Geneva",
      year          = "2023",
      url           = "https://cds.cern.ch/record/2866276",
}

@article{fastjet,
    author = "Cacciari, Matteo and Salam, Gavin P. and Soyez, Gregory",
    title = "{FastJet User Manual}",
    eprint = "1111.6097",
    archivePrefix = "arXiv",
    primaryClass = "hep-ph",
    reportNumber = "CERN-PH-TH-2011-297",
    doi = "10.1140/epjc/s10052-012-1896-2",
    journal = "Eur. Phys. J. C",
    volume = "72",
    pages = "1896",
    year = "2012"
}

@article{ghost_association,
    author = "Cacciari, Matteo and Salam, Gavin P. and Soyez, Gregory",
    title = "{The Catchment Area of Jets}",
    eprint = "0802.1188",
    archivePrefix = "arXiv",
    primaryClass = "hep-ph",
    reportNumber = "LPTHE-07-02",
    doi = "10.1088/1126-6708/2008/04/005",
    journal = "JHEP",
    volume = "04",
    pages = "005",
    year = "2008"
}

@article{kt_splitting,
    author = "Thaler, Jesse and Wang, Lian-Tao",
    title = "{Strategies to Identify Boosted Tops}",
    eprint = "0806.0023",
    archivePrefix = "arXiv",
    primaryClass = "hep-ph",
    doi = "10.1088/1126-6708/2008/07/092",
    journal = "JHEP",
    volume = "07",
    pages = "092",
    year = "2008"
}

@article{thrusts,
    author = "Chen, Chunhui",
    title = "{New approach to identifying boosted hadronically decaying particles using jet substructure in its center-of-mass frame}",
    eprint = "1112.2567",
    archivePrefix = "arXiv",
    primaryClass = "hep-ph",
    doi = "10.1103/PhysRevD.85.034007",
    journal = "Phys. Rev. D",
    volume = "85",
    pages = "034007",
    year = "2012"
}

@article{mlp,
  title={Learning representations by back-propagating errors},
  author={Rumelhart, David E and Hinton, Geoffrey E and Williams, Ronald J},
  journal={Nature},
  volume={323},
  number={6088},
  pages={533--536},
  year={1986},
  publisher={Nature Publishing Group},
  doi="10.1038/323533a0"
}

@article{cnn3,
    author = "de Oliveira, Luke and Kagan, Michael and Mackey, Lester and Nachman, Benjamin and Schwartzman, Ariel",
    title = "{Jet-images \textemdash{} deep learning edition}",
    eprint = "1511.05190",
    archivePrefix = "arXiv",
    primaryClass = "hep-ph",
    doi = "10.1007/JHEP07(2016)069",
    journal = "JHEP",
    volume = "07",
    pages = "069",
    year = "2016"
}

@article{cnn4,
    author = "Macaluso, Sebastian and Shih, David",
    title = "{Pulling Out All the Tops with Computer Vision and Deep Learning}",
    eprint = "1803.00107",
    archivePrefix = "arXiv",
    primaryClass = "hep-ph",
    doi = "10.1007/JHEP10(2018)121",
    journal = "JHEP",
    volume = "10",
    pages = "121",
    year = "2018"
}

@article{cnn5,
    author = "Kasieczka, Gregor and Plehn, Tilman and Russell, Michael and Schell, Torben",
    title = "{Deep-learning Top Taggers or The End of QCD?}",
    eprint = "1701.08784",
    archivePrefix = "arXiv",
    primaryClass = "hep-ph",
    reportNumber = "MCNET-17-07",
    doi = "10.1007/JHEP05(2017)006",
    journal = "JHEP",
    volume = "05",
    pages = "006",
    year = "2017"
}

@article{pythia_weights,
    author = "Mrenna, S. and Skands, P.",
    title = "{Automated Parton-Shower Variations in Pythia 8}",
    eprint = "1605.08352",
    archivePrefix = "arXiv",
    primaryClass = "hep-ph",
    reportNumber = "COEPP-MN-16-13, MCNET-16-14, FERMILAB-PUB-16-186-CD, MCNET-16-15",
    doi = "10.1103/PhysRevD.94.074005",
    journal = "Phys. Rev. D",
    volume = "94",
    number = "7",
    pages = "074005",
    year = "2016"
}

@article{class_party,
    author = "Qu, Huilin and Li, Congqiao and Qian, Sitian",
    title = "{Particle Transformer for Jet Tagging}",
    eprint = "2202.03772",
    archivePrefix = "arXiv",
    primaryClass = "hep-ph",
    month = "2",
    year = "2024"
}

@article{pelican1,
    author = "Bogatskiy, Alexander and Hoffman, Timothy and Miller, David W. and Offermann, Jan T.",
    title = "{PELICAN: Permutation Equivariant and Lorentz Invariant or Covariant Aggregator Network for Particle Physics}",
    eprint = "2211.00454",
    archivePrefix = "arXiv",
    primaryClass = "hep-ph",
    month = "11",
    year = "2022"
}

@article{pelican2,
    author = "Bogatskiy, Alexander and Hoffman, Timothy and Miller, David W. and Offermann, Jan T. and Liu, Xiaoyang",
    title = "{Explainable equivariant neural networks for particle physics: PELICAN}",
    eprint = "2307.16506",
    archivePrefix = "arXiv",
    primaryClass = "hep-ph",
    doi = "10.1007/JHEP03(2024)113",
    journal = "JHEP",
    volume = "03",
    pages = "113",
    year = "2024"
}

@article{pelican3,
    author = "Bogatskiy, Alexander and Hoffman, Timothy and Offermann, Jan T.",
    title = "{19 Parameters Is All You Need: Tiny Neural Networks for Particle Physics}",
    journal = "{Advances in neural information processing systems}",
    eprint = "2310.16121",
    archivePrefix = "arXiv",
    primaryClass = "hep-ph",
    month = "10",
    year = "2023"
}

@article{deepsets,
  title={Deep sets},
  author={Zaheer, Manzil and Kottur, Satwik and Ravanbakhsh, Siamak and Poczos, Barnabas and Salakhutdinov, Russ R and Smola, Alexander J},
  journal={Advances in neural information processing systems},
  eprint={1703.06114},
  archivePrefix={arXiv},
  primaryClass={cs.LG},
  year={2018}
}

@article{lorentznet1,
    author = "Gong, Shiqi and Meng, Qi and Zhang, Jue and Qu, Huilin and Li, Congqiao and Qian, Sitian and Du, Weitao and Ma, Zhi-Ming and Liu, Tie-Yan",
    title = "{An efficient Lorentz equivariant graph neural network for jet tagging}",
    eprint = "2201.08187",
    archivePrefix = "arXiv",
    primaryClass = "hep-ph",
    doi = "10.1007/JHEP07(2022)030",
    journal = "JHEP",
    volume = "07",
    pages = "030",
    year = "2022"
}

@article{lorentznet2,
    author = "Li, Congqiao and Qu, Huilin and Qian, Sitian and Meng, Qi and Gong, Shiqi and Zhang, Jue and Liu, Tie-Yan and Li, Qiang",
    title = "{Does Lorentz-symmetric design boost network performance in jet physics?}",
    eprint = "2208.07814",
    archivePrefix = "arXiv",
    primaryClass = "hep-ph",
    doi = "10.1103/PhysRevD.109.056003",
    journal = "Phys. Rev. D",
    volume = "109",
    number = "5",
    pages = "056003",
    year = "2024"
}

@article{isnotallyouneed,
    author = "Bright-Thonney, Samuel and Nachman, Benjamin and Thaler, Jesse",
    title = "{Safe but Incalculable: Energy-weighting is not all you need}",
    eprint = "2311.07652",
    archivePrefix = "arXiv",
    primaryClass = "hep-ph",
    reportNumber = "MIT-CTP 5641",
    month = "11",
    year = "2024"
}

@article{therightway,
    author = "Dolan, Matthew J. and Gargalionis, John and Ore, Ayodele",
    title = "{Quark-versus-gluon tagging in CMS Open Data with CWoLa and TopicFlow}",
    eprint = "2312.03434",
    archivePrefix = "arXiv",
    primaryClass = "hep-ph",
    month = "12",
    year = "2023"
}

@article{sherpa,
    author = "Bothmann, Enrico and others",
    collaboration = "Sherpa",
    title = "{Event Generation with Sherpa 2.2}",
    eprint = "1905.09127",
    archivePrefix = "arXiv",
    primaryClass = "hep-ph",
    reportNumber = "FERMILAB-PUB-19-218-T, SLAC-PUB-17433, IPPP/19/42, MCNET-19-11",
    doi = "10.21468/SciPostPhys.7.3.034",
    journal = "SciPost Phys.",
    volume = "7",
    number = "3",
    pages = "034",
    year = "2019"
}

@article{heavy_vector_bosons,
    author = "Altarelli, Guido and Mele, B. and Ruiz-Altaba, M.",
    title = "{Searching for New Heavy Vector Bosons in $p \bar{p}$ Colliders}",
    reportNumber = "CERN-TH-5323-89",
    doi = "10.1007/BF01556677",
    journal = "Z. Phys. C",
    volume = "45",
    pages = "109",
    year = "1989",
    note = "[Erratum: Z.Phys.C 47, 676 (1990)]"
}

@article{boost2011,
    author = "Altheimer, A. and others",
    title = "{Jet Substructure at the Tevatron and LHC: New results, new tools, new benchmarks}",
    eprint = "1201.0008",
    archivePrefix = "arXiv",
    primaryClass = "hep-ph",
    reportNumber = "SLAC-R-990, FERMILAB-PUB-12-897-T",
    doi = "10.1088/0954-3899/39/6/063001",
    journal = "J. Phys. G",
    volume = "39",
    pages = "063001",
    year = "2012"
}

@article{boost2010,
    author = "Abdesselam, A. and others",
    title = "{Boosted Objects: A Probe of Beyond the Standard Model Physics}",
    eprint = "1012.5412",
    archivePrefix = "arXiv",
    primaryClass = "hep-ph",
    reportNumber = "SLAC-PUB-15081, FERMILAB-PUB-10-617-CMS",
    doi = "10.1140/epjc/s10052-011-1661-y",
    journal = "Eur. Phys. J. C",
    volume = "71",
    pages = "1661",
    year = "2011"
}

@article{jss_ml_1,
    author = "Larkoski, Andrew J. and Moult, Ian and Nachman, Benjamin",
    title = "{Jet Substructure at the Large Hadron Collider: A Review of Recent Advances in Theory and Machine Learning}",
    eprint = "1709.04464",
    archivePrefix = "arXiv",
    primaryClass = "hep-ph",
    doi = "10.1016/j.physrep.2019.11.001",
    journal = "Phys. Rept.",
    volume = "841",
    pages = "1--63",
    year = "2020"
}

@article{jss_ml_2,
    author = "Guest, Dan and Cranmer, Kyle and Whiteson, Daniel",
    title = "{Deep Learning and its Application to LHC Physics}",
    eprint = "1806.11484",
    archivePrefix = "arXiv",
    primaryClass = "hep-ex",
    doi = "10.1146/annurev-nucl-101917-021019",
    journal = "Ann. Rev. Nucl. Part. Sci.",
    volume = "68",
    pages = "161--181",
    year = "2018"
}

@article{STDM-2024-01,
    author = "{ATLAS Collaboration}",
    title = "{Measurements of Lund subjet multiplicities in 13 TeV proton-proton collisions with the ATLAS detector}",
    eprint = "2402.13052",
    archivePrefix = "arXiv",
    primaryClass = "hep-ex",
    reportNumber = "CERN-EP-2024-029",
    month = "2",
    year = "2024"
}

@article{dnn_tagger,
    author = "Pearkes, Jannicke and Fedorko, Wojciech and Lister, Alison and Gay, Colin",
    title = "{Jet Constituents for Deep Neural Network Based Top Quark Tagging}",
    eprint = "1704.02124",
    archivePrefix = "arXiv",
    primaryClass = "hep-ex",
    month = "4",
    year = "2017"
}

@article{uncertainty_aware,
    author = "Ghosh, Aishik and Nachman, Benjamin and Whiteson, Daniel",
    title = "{Uncertainty-aware machine learning for high energy physics}",
    eprint = "2105.08742",
    archivePrefix = "arXiv",
    primaryClass = "physics.data-an",
    doi = "10.1103/PhysRevD.104.056026",
    journal = "Phys. Rev. D",
    volume = "104",
    number = "5",
    pages = "056026",
    year = "2021"
}

@article{cautionary_tale,
    author = "Ghosh, Aishik and Nachman, Benjamin",
    title = "{A cautionary tale of decorrelating theory uncertainties}",
    eprint = "2109.08159",
    archivePrefix = "arXiv",
    primaryClass = "hep-ph",
    doi = "10.1140/epjc/s10052-022-10012-w",
    journal = "Eur. Phys. J. C",
    volume = "82",
    number = "1",
    pages = "46",
    year = "2022"
}

@article{adversarial_training,
    author = "Stein, Annika and Coubez, Xavier and Mondal, Spandan and Novak, Andrzej and Schmidt, Alexander",
    title = "{Improving Robustness of Jet Tagging Algorithms with Adversarial Training}",
    eprint = "2203.13890",
    archivePrefix = "arXiv",
    primaryClass = "physics.data-an",
    doi = "10.1007/s41781-022-00087-1",
    journal = "Comput. Softw. Big Sci.",
    volume = "6",
    number = "1",
    pages = "15",
    year = "2022"
}

@article{perf_vs_resilience,
    author = "Butter, Anja and Dillon, Barry M. and Plehn, Tilman and Vogel, Lorenz",
    title = "{Performance versus resilience in modern quark-gluon tagging}",
    eprint = "2212.10493",
    archivePrefix = "arXiv",
    primaryClass = "hep-ph",
    doi = "10.21468/SciPostPhysCore.6.4.085",
    journal = "SciPost Phys. Core",
    volume = "6",
    pages = "085",
    year = "2023"
}

@article{cwola,
    author = "Metodiev, Eric M. and Nachman, Benjamin and Thaler, Jesse",
    title = "{Classification without labels: Learning from mixed samples in high energy physics}",
    eprint = "1708.02949",
    archivePrefix = "arXiv",
    primaryClass = "hep-ph",
    reportNumber = "MIT--CTP-4922",
    doi = "10.1007/JHEP10(2017)174",
    journal = "JHEP",
    volume = "10",
    pages = "174",
    year = "2017"
}

@inbook{auc,
	address = {New York, NY},
	author = {Melo, Francisco},
	booktitle = {Encyclopedia of Systems Biology},
	doi = {10.1007/978-1-4419-9863-7_209},
	editor = {Dubitzky, Werner and Wolkenhauer, Olaf and Cho, Kwang-Hyun and Yokota, Hiroki},
	isbn = {978-1-4419-9863-7},
	pages = {38--39},
	publisher = {Springer New York},
	title = {Area under the ROC Curve},
	year = {2013},
}

@misc{open_data,
    author = "{ATLAS Collaboration}",
    title = "{ATLAS top tagging open data set with systematic uncertainties}",
    url = "http://doi.org/10.7483/OPENDATA.ATLAS.SOAY.LABE"
}

@Article{Evans:2008zzb,
      author         = "Evans, Lyndon and Bryant, Philip",
      title          = "{LHC Machine}",
      journal        = "JINST",
      volume         = "3",
      pages          = "S08001",
      doi            = "10.1088/1748-0221/3/08/S08001",
      year           = "2008",
      SLACcitation   = "%%CITATION = JINST,3,S08001;%%",
}

@article{Ball:2012cx,
      author         = "{NNPDF Collaboration} and Ball, Richard D. and others",
      title          = "{Parton distributions with LHC data}",
      journal        = "Nucl. Phys. B",
      volume         = "867",
      year           = "2013",
      pages          = "244",
      doi            = "10.1016/j.nuclphysb.2012.10.003",
      eprint         = "1207.1303",
      archivePrefix  = "arXiv",
      primaryClass   = "hep-ph",
      reportNumber   = "EDINBURGH-2012-08, IFUM-FT-997, FR-PHENO-2012-014,
                        RWTH-TTK-12-25, CERN-PH-TH-2012-037, SFB-CPP-12-47\,
                        --CERN-PH-TH-2012-037",
      SLACcitation   = "%%CITATION = ARXIV:1207.1303;%%"
}

@article{Ball:2014uwa,
      author         = "{NNPDF Collaboration} and Ball, Richard D. and others",
      title          = "{Parton distributions for the LHC run II}",
      journal        = "JHEP",
      volume         = "04",
      year           = "2015",
      pages          = "040",
      doi            = "10.1007/JHEP04(2015)040",
      eprint         = "1410.8849",
      archivePrefix  = "arXiv",
      primaryClass   = "hep-ph",
      reportNumber   = "EDINBURGH-2014-15, IFUM-1034-FT, CERN-PH-TH-2013-253,
                        OUTP-14-11P, CAVENDISH-HEP-14-11",
      SLACcitation   = "%%CITATION = ARXIV:1410.8849;%%"
}

@Article{Sjostrand:2006za,
     author    = "Sj{\"o}strand, Torbjorn and Mrenna, Stephen and Skands, Peter",
     title     = "{PYTHIA 6.4 physics and manual}",
     journal   = "JHEP",
     volume    = "05",
     year      = "2006",
     pages     = "026",
     eprint    = "hep-ph/0603175",
     archivePrefix = "arXiv",
     doi       = "10.1088/1126-6708/2006/05/026",
     SLACcitation  = "%%CITATION = HEP-PH/0603175;%%"
}

@Article{Lange:2001uf,
      author         = "Lange, D. J.",
      title          = "{The EvtGen particle decay simulation package}",
      booktitle      = "{Proceedings, 7th International Conference on B physics
                        at hadron machines (BEAUTY 2000)}",
      journal        = "Nucl. Instrum. Meth. A",
      volume         = "462",
      year           = "2001",
      pages          = "152",
      doi            = "10.1016/S0168-9002(01)00089-4",
      SLACcitation   = "%%CITATION = NUIMA,A462,152;%%"
}

@Article{Nason:2004rx,
      author         = "Nason, Paolo",
      title          = "{A new method for combining NLO QCD with shower Monte Carlo algorithms}",
      journal        = "JHEP",
      volume         = "11",
      pages          = "040",
      doi            = "10.1088/1126-6708/2004/11/040",
      year           = "2004",
      eprint         = "hep-ph/0409146",
      archivePrefix  = "arXiv",
}

@Article{Frixione:2007vw,
      author         = "Frixione, Stefano and Nason, Paolo and Oleari, Carlo",
      title          = "{Matching NLO QCD computations with parton shower
                        simulations: the POWHEG method}",
      journal        = "JHEP",
      volume         = "11",
      pages          = "070",
      doi            = "10.1088/1126-6708/2007/11/070",
      year           = "2007",
      eprint         = "0709.2092",
      archivePrefix  = "arXiv",
      primaryClass   = "hep-ph",
}

@Article{Alioli:2010xd,
      author         = "Alioli, Simone and Nason, Paolo and Oleari, Carlo and Re,
                        Emanuele",
      title          = "{A general framework for implementing NLO calculations in
                        shower Monte Carlo programs: the POWHEG BOX}",
      journal        = "JHEP",
      volume         = "06",
      pages          = "043",
      doi            = "10.1007/JHEP06(2010)043",
      year           = "2010",
      eprint         = "1002.2581",
      archivePrefix  = "arXiv",
      primaryClass   = "hep-ph",
}

@article{Frixione:2007nw,
      author         = "Frixione, Stefano and Ridolfi, Giovanni and Nason, Paolo",
      title          = "{A positive-weight next-to-leading-order Monte Carlo for
                        heavy flavour hadroproduction}",
      journal        = "JHEP",
      volume         = "09",
      pages          = "126",
      doi            = "10.1088/1126-6708/2007/09/126",
      year           = "2007",
      eprint         = "0707.3088",
      archivePrefix  = "arXiv",
      primaryClass   = "hep-ph",
      SLACcitation   = "%%CITATION = ARXIV:0707.3088;%%",
}

@article{Winter:2003tt,
      author         = "Winter, Jan-Christopher and Krauss, Frank and Soff,
                        Gerhard",
      title          = "{A modified cluster-hadronisation model}",
      journal        = "Eur. Phys. J. C",
      volume         = "36",
      year           = "2004",
      pages          = "381-395",
      doi            = "10.1140/epjc/s2004-01960-8",
      eprint         = "hep-ph/0311085",
      archivePrefix  = "arXiv",
      XprimaryClass   = "hep-ph",
      reportNumber   = "CERN-TH-2003-272",
      SLACcitation   = "%%CITATION = HEP-PH/0311085;%%"
}

@article{Agostinelli:2002hh,
      author         = "Agostinelli, S. and others",
      title          = "{\textsc{Geant4} -- a simulation toolkit}",
      journal        = "Nucl. Instrum. Meth. A",
      volume         = "506",
      year           = "2003",
      pages          = "250",
      doi            = "10.1016/S0168-9002(03)01368-8",
      reportNumber   = "SLAC-PUB-9350, FERMILAB-PUB-03-339",
      SLACcitation   = "%%CITATION = NUIMA,A506,250;%%"
}

@Article{PERF-2007-01,
    author         = "{ATLAS Collaboration}",
    title          = "{The ATLAS Experiment at the CERN Large Hadron Collider}",
    journal        = "JINST",
    volume         = "3",
    year           = "2008",
    pages          = "S08003",
    doi            = "10.1088/1748-0221/3/08/S08003",
    primaryClass   = "hep-ex",
}

@Article{SOFT-2010-01,
    author         = "{ATLAS Collaboration}",
    title          = "{The ATLAS Simulation Infrastructure}",
    journal        = "Eur. Phys. J. C",
    volume         = "70",
    year           = "2010",
    pages          = "823",
    doi            = "10.1140/epjc/s10052-010-1429-9",
    eprint         = "1005.4568",
    archivePrefix  = "arXiv",
    primaryClass   = "physics.ins-det",
}

@Article{STDM-2012-12,
    author         = "{ATLAS Collaboration}",
    title          = "{Measurement of \(k_{\text{T}}\) splitting scales in \(W \rightarrow \ell\nu\) events at \(\sqrt{s} = 7\,\text{TeV}\) with the ATLAS detector}",
    journal        = "Eur. Phys. J. C",
    volume         = "73",
    year           = "2013",
    pages          = "2432",
    doi            = "10.1140/epjc/s10052-013-2432-8",
    reportNumber   = "CERN-PH-EP-2013-003",
    eprint         = "1302.1415",
    archivePrefix  = "arXiv",
    primaryClass   = "hep-ex",
}

@Article{PERF-2014-07,
    author         = "{ATLAS Collaboration}",
    title          = "{Topological cell clustering in the ATLAS calorimeters and its performance in LHC Run~1}",
    journal        = "Eur. Phys. J. C",
    volume         = "77",
    year           = "2017",
    pages          = "490",
    doi            = "10.1140/epjc/s10052-017-5004-5",
    reportNumber   = "CERN-PH-EP-2015-304",
    eprint         = "1603.02934",
    archivePrefix  = "arXiv",
    primaryClass   = "hep-ex",
}

@Article{PERF-2015-04,
    author         = "{ATLAS Collaboration}",
    title          = "{Identification of high transverse momentum top quarks in \(pp\) collisions at \(\sqrt{s} = 8\,\text{TeV}\) with the ATLAS detector}",
    journal        = "JHEP",
    volume         = "06",
    year           = "2016",
    pages          = "093",
    doi            = "10.1007/JHEP06(2016)093",
    reportNumber   = "CERN-EP-2016-010",
    eprint         = "1603.03127",
    archivePrefix  = "arXiv",
    primaryClass   = "hep-ex",
}

@Article{PERF-2015-05,
    author         = "{ATLAS Collaboration}",
    title          = "{A measurement of the calorimeter response to single hadrons and determination of the jet energy scale uncertainty using LHC Run-1 \(pp\)-collision data with the ATLAS detector}",
    journal        = "Eur. Phys. J. C",
    volume         = "77",
    year           = "2017",
    pages          = "26",
    doi            = "10.1140/epjc/s10052-016-4580-0",
    reportNumber   = "CERN-EP-2016-149",
    eprint         = "1607.08842",
    archivePrefix  = "arXiv",
    primaryClass   = "hep-ex",
}

@Article{PERF-2015-08,
    author         = "{ATLAS Collaboration}",
    title          = "{Performance of the ATLAS track reconstruction algorithms in dense environments in LHC Run~2}",
    journal        = "Eur. Phys. J. C",
    volume         = "77",
    year           = "2017",
    pages          = "673",
    doi            = "10.1140/epjc/s10052-017-5225-7",
    reportNumber   = "CERN-EP-2017-045",
    eprint         = "1704.07983",
    archivePrefix  = "arXiv",
    primaryClass   = "hep-ex",
}

@Article{PERF-2015-09,
    author         = "{ATLAS Collaboration}",
    title          = "{Jet reconstruction and performance using particle flow with the ATLAS Detector}",
    journal        = "Eur. Phys. J. C",
    volume         = "77",
    year           = "2017",
    pages          = "466",
    doi            = "10.1140/epjc/s10052-017-5031-2",
    reportNumber   = "CERN-EP-2017-024",
    eprint         = "1703.10485",
    archivePrefix  = "arXiv",
    primaryClass   = "hep-ex",
}

@Article{TOPQ-2015-10,
    author         = "{ATLAS Collaboration}",
    title          = "{Measurement of the charge asymmetry in highly boosted top-quark pair production in \(\sqrt{s} = 8\,\text{TeV}\) \(pp\) collision data collected by the ATLAS experiment}",
    journal        = "Phys. Lett. B",
    volume         = "756",
    year           = "2016",
    pages          = "52",
    doi            = "10.1016/j.physletb.2016.02.055",
    reportNumber   = "CERN-PH-EP-2015-303",
    eprint         = "1512.06092",
    archivePrefix  = "arXiv",
    primaryClass   = "hep-ex",
}

@Article{STDM-2017-04,
    author         = "{ATLAS Collaboration}",
    title          = "{Measurement of the Soft-Drop Jet Mass in \(pp\) Collisions at \(\sqrt{s} = 13\,\text{TeV}\) with the ATLAS detector}",
    journal        = "Phys. Rev. Lett.",
    volume         = "121",
    year           = "2018",
    pages          = "092001",
    doi            = "10.1103/PhysRevLett.121.092001",
    reportNumber   = "CERN-EP-2017-231",
    eprint         = "1711.08341",
    archivePrefix  = "arXiv",
    primaryClass   = "hep-ex",
}

@Article{STDM-2017-33,
    author         = "{ATLAS Collaboration}",
    title          = "{Measurement of soft-drop jet observables in \(pp\) collisions with the ATLAS detector at \(\sqrt{s} = 13\,\text{TeV}\)}",
    journal        = "Phys. Rev. D",
    volume         = "101",
    year           = "2020",
    pages          = "052007",
    doi            = "10.1103/PhysRevD.101.052007",
    reportNumber   = "CERN-EP-2019-269",
    eprint         = "1912.09837",
    archivePrefix  = "arXiv",
    primaryClass   = "hep-ex",
}

@Article{STDM-2017-34,
    author         = "{ATLAS Collaboration}",
    title          = "{Measurement of jet-substructure observables in top quark, \(W\) boson and light jet production in proton--proton collisions at \(\sqrt{s} = 13\,\text{TeV}\) with the ATLAS detector}",
    journal        = "JHEP",
    volume         = "08",
    year           = "2019",
    pages          = "033",
    doi            = "10.1007/JHEP08(2019)033",
    reportNumber   = "CERN-EP-2019-011",
    eprint         = "1903.02942",
    archivePrefix  = "arXiv",
    primaryClass   = "hep-ex",
}

@Article{EXOT-2018-48,
    author         = "{ATLAS Collaboration}",
    title          = "{Search for  \(t\bar{t}\) resonances in fully hadronic final states in \(pp\) collisions at \(\sqrt{s} = 13\,\text{TeV}\) with the ATLAS detector}",
    journal        = "JHEP",
    volume         = "10",
    year           = "2020",
    pages          = "061",
    doi            = "10.1007/JHEP10(2020)061",
    reportNumber   = "CERN-EP-2020-055",
    eprint         = "2005.05138",
    archivePrefix  = "arXiv",
    primaryClass   = "hep-ex",
}

@Article{JETM-2018-03,
    author         = "{ATLAS Collaboration}",
    title          = "{Performance of top-quark and \(W\)-boson tagging with ATLAS in Run~2 of the LHC}",
    journal        = "Eur. Phys. J. C",
    volume         = "79",
    year           = "2019",
    pages          = "375",
    doi            = "10.1140/epjc/s10052-019-6847-8",
    reportNumber   = "CERN-EP-2018-192",
    eprint         = "1808.07858",
    archivePrefix  = "arXiv",
    primaryClass   = "hep-ex",
}

@Article{JETM-2018-06,
    author         = "{ATLAS Collaboration}",
    title          = "{Optimisation of large-radius jet reconstruction for the ATLAS detector in \(13\,\text{TeV}\) proton--proton collisions}",
    journal        = "Eur. Phys. J. C",
    volume         = "81",
    year           = "2021",
    pages          = "334",
    doi            = "10.1140/epjc/s10052-021-09054-3",
    reportNumber   = "CERN-EP-2020-134",
    eprint         = "2009.04986",
    archivePrefix  = "arXiv",
    primaryClass   = "hep-ex",
}

@Article{JETM-2018-08,
    author         = "{ATLAS Collaboration}",
    title          = "{Measurement of the energy response of the ATLAS calorimeter to charged pions from \(W^{\pm}\to\tau^{\pm}(\to\pi^{\pm}\nu_{\tau})\nu_{\tau}\) events in Run~2 data}",
    journal        = "Eur. Phys. J. C",
    volume         = "82",
    year           = "2022",
    pages          = "223",
    doi            = "10.1140/epjc/s10052-022-10117-2",
    reportNumber   = "CERN-EP-2021-147",
    eprint         = "2108.09043",
    archivePrefix  = "arXiv",
    primaryClass   = "hep-ex",
}

@Article{STDM-2018-57,
    author         = "{ATLAS Collaboration}",
    title          = "{Measurement of the Lund Jet Plane Using Charged Particles in \(13\,\text{TeV}\) Proton--Proton Collisions with the ATLAS Detector}",
    journal        = "Phys. Rev. Lett.",
    volume         = "124",
    year           = "2020",
    pages          = "222002",
    doi            = "10.1103/PhysRevLett.124.222002",
    reportNumber   = "CERN-EP-2020-030",
    eprint         = "2004.03540",
    archivePrefix  = "arXiv",
    primaryClass   = "hep-ex",
}

@Article{TOPQ-2018-11,
    author         = "{ATLAS Collaboration}",
    title          = "{Differential \(t\bar{t}\) cross-section measurements using boosted top quarks in the all-hadronic final state with \(139\,\text{fb}^{-1}\) of ATLAS data}",
    journal        = "JHEP",
    volume         = "04",
    year           = "2023",
    pages          = "080",
    doi            = "10.1007/JHEP04(2023)080",
    reportNumber   = "CERN-EP-2022-026",
    eprint         = "2205.02817",
    archivePrefix  = "arXiv",
    primaryClass   = "hep-ex",
}

@Article{EXOT-2019-07,
    author         = "{ATLAS Collaboration}",
    title          = "{Search for single production of a vectorlike \(T\) quark decaying into a Higgs boson and top quark with fully hadronic final states using the ATLAS detector}",
    journal        = "Phys. Rev. D",
    volume         = "105",
    year           = "2022",
    pages          = "092012",
    doi            = "10.1103/PhysRevD.105.092012",
    reportNumber   = "CERN-EP-2021-249",
    eprint         = "2201.07045",
    archivePrefix  = "arXiv",
    primaryClass   = "hep-ex",
}

@Article{IDTR-2019-05,
    author         = "{ATLAS Collaboration}",
    title          = "{Alignment of the ATLAS Inner Detector in Run 2}",
    journal        = "Eur. Phys. J. C",
    volume         = "80",
    year           = "2020",
    pages          = "1194",
    doi            = "10.1140/epjc/s10052-020-08700-6",
    reportNumber   = "CERN-EP-2020-108",
    eprint         = "2007.07624",
    archivePrefix  = "arXiv",
    primaryClass   = "hep-ex",
}

@Article{TOPQ-2019-22,
    author         = "{ATLAS Collaboration}",
    title          = "{Measurement of jet substructure in boosted \(t\bar{t}\) events with the ATLAS detector using \(140\,\text{fb}^{-1}\) of \(13\,\text{TeV}\) \(pp\) collisions}",
    year           = "2023",
    reportNumber   = "CERN-EP-2023-239",
    eprint         = "2312.03797",
    archivePrefix  = "arXiv",
    primaryClass   = "hep-ex",
}

@Article{TOPQ-2019-28,
    author         = "{ATLAS Collaboration}",
    title          = "{Measurement of the energy asymmetry in \(t\bar{t}j\) production at \(13\,\text{TeV}\) with the ATLAS experiment and interpretation in the SMEFT framework}",
    journal        = "Eur. Phys. J. C",
    volume         = "82",
    year           = "2022",
    pages          = "374",
    doi            = "10.1140/epjc/s10052-022-10101-w",
    reportNumber   = "CERN-EP-2021-181",
    eprint         = "2110.05453",
    archivePrefix  = "arXiv",
    primaryClass   = "hep-ex",
}

@Article{DAPR-2021-01,
    author         = "{ATLAS Collaboration}",
    title          = "{Luminosity determination in \(pp\) collisions at \(\sqrt{s} = 13\,\text{TeV}\) using the ATLAS detector at the LHC}",
    journal        = "Eur. Phys. J. C",
    volume         = "83",
    year           = "2023",
    pages          = "982",
    doi            = "10.1140/epjc/s10052-023-11747-w",
    reportNumber   = "CERN-EP-2022-281",
    eprint         = "2212.09379",
    archivePrefix  = "arXiv",
    primaryClass   = "hep-ex",
}

@Article{SOFT-2022-02,
    author         = "{ATLAS Collaboration}",
    title          = "{Software and computing for Run~3 of the ATLAS experiment at the LHC}",
    year           = "2024",
    reportNumber   = "CERN-EP-2024-100",
    eprint         = "2404.06335",
    archivePrefix  = "arXiv",
    primaryClass   = "hep-ex",
}

@Booklet{ATL-SOFT-PUB-2023-001,
    author         = "{ATLAS Collaboration}",
    title          = "{ATLAS Computing Acknowledgements}",
    howpublished   = "{ATL-SOFT-PUB-2023-001}",
    url            = "https://cds.cern.ch/record/2869272",
    year           = "2023",
}

@Article{LUCID2,
    author         = "Avoni, G. and others",
    title          = "{The new LUCID-2 detector for luminosity measurement and monitoring in ATLAS}",
    journal        = "JINST",
    volume         = "13",
    year           = "2018",
    pages          = "P07017",
    doi            = "10.1088/1748-0221/13/07/P07017",
}

@Article{CMS-CMS-00-001,
    author         = "{CMS Collaboration}",
    title          = "{The CMS Experiment at the CERN LHC}",
    journal        = "JINST",
    volume         = "3",
    year           = "2008",
    pages          = "S08004",
    doi            = "10.1088/1748-0221/3/08/S08004",
}

@Article{CMS-TOP-15-015,
    author         = "{CMS Collaboration}",
    title          = "{Measurement of the jet mass in highly boosted \(t\bar{t}\) events from \(pp\) collisions at \(\sqrt{s} = 8\,\text{TeV}\)}",
    journal        = "Eur. Phys. J. C",
    volume         = "77",
    year           = "2017",
    pages          = "467",
    doi            = "10.1140/epjc/s10052-017-5030-3",
    reportNumber   = "CERN-EP-2017-018",
    eprint         = "1703.06330",
    archivePrefix  = "arXiv",
    primaryClass   = "hep-ex",
}

@Article{CMS-B2G-16-005,
    author         = "{CMS Collaboration}",
    title          = "{Search for electroweak production of a vector-like quark decaying to a top quark and a Higgs boson using boosted topologies in fully hadronic final states}",
    journal        = "JHEP",
    volume         = "04",
    year           = "2017",
    pages          = "136",
    doi            = "10.1007/JHEP04(2017)136",
    reportNumber   = "CERN-EP-2016-290",
    eprint         = "1612.05336",
    archivePrefix  = "arXiv",
    primaryClass   = "hep-ex",
}

@Article{CMS-JME-18-002,
    author         = "{CMS Collaboration}",
    title          = "{Identification of heavy, energetic, hadronically decaying particles using machine-learning techniques}",
    journal        = "JINST",
    volume         = "15",
    year           = "2020",
    pages          = "P06005",
    doi            = "10.1088/1748-0221/15/06/P06005",
    reportNumber   = "CERN-EP-2020-037",
    eprint         = "2004.08262",
    archivePrefix  = "arXiv",
    primaryClass   = "hep-ex",
}

@Article{CMS-TOP-18-011,
    author         = "{CMS Collaboration}",
    title          = "{Measurement of the \(t\bar{t} b\bar{b}\) production cross section in the all-jet final state in \(pp\) collisions at \(\sqrt{s} = 13\,\text{TeV}\)}",
    journal        = "Phys. Lett. B",
    volume         = "803",
    year           = "2020",
    pages          = "135285",
    doi            = "10.1016/j.physletb.2020.135285",
    reportNumber   = "CERN-EP-2019-183",
    eprint         = "1909.05306",
    archivePrefix  = "arXiv",
    primaryClass   = "hep-ex",
}

@Article{CMS-TOP-18-013,
    author         = "{CMS Collaboration}",
    title          = "{Measurement of differential \(t\bar{t}\) production cross sections using top quarks at large transverse momenta in \(pp\) collisions at \(\sqrt{s} = 13\,\text{TeV}\)}",
    journal        = "Phys. Rev. D",
    volume         = "103",
    year           = "2021",
    pages          = "052008",
    doi            = "10.1103/PhysRevD.103.052008",
    reportNumber   = "CERN-EP-2020-121",
    eprint         = "2008.07860",
    archivePrefix  = "arXiv",
    primaryClass   = "hep-ex",
}

@Article{CMS-B2G-19-003,
    author         = "{CMS Collaboration}",
    title          = "{Search for a heavy resonance decaying to a top quark and a \(W\) boson at \(\sqrt{s} = 13\,\text{TeV}\) in the fully hadronic final state}",
    journal        = "JHEP",
    volume         = "12",
    year           = "2021",
    pages          = "106",
    doi            = "10.1007/JHEP12(2021)106",
    reportNumber   = "CERN-EP-2021-044",
    eprint         = "2104.12853",
    archivePrefix  = "arXiv",
    primaryClass   = "hep-ex",
}

@Article{CMS-B2G-19-004,
    author         = "{CMS Collaboration}",
    title          = "{Search for single production of a vector-like \(T\) quark decaying to a top quark and a \(Z\) boson in the final state with jets and missing transverse momentum at \(\sqrt{s} = 13\,\text{TeV}\)}",
    journal        = "JHEP",
    volume         = "05",
    year           = "2022",
    pages          = "093",
    doi            = "10.1007/JHEP05(2022)093",
    reportNumber   = "CERN-EP-2021-115",
    eprint         = "2201.02227",
    archivePrefix  = "arXiv",
    primaryClass   = "hep-ex",
}

@Article{CMS-B2G-20-002,
    author         = "{CMS Collaboration}",
    title          = "{Search for a \(W'\) boson decaying to a vector-like quark and a top or bottom quark in the all-jets final state at \(\sqrt{s} = 13\,\text{TeV}\)}",
    journal        = "JHEP",
    volume         = "09",
    year           = "2022",
    pages          = "088",
    doi            = "10.1007/JHEP09(2022)088",
    reportNumber   = "CERN-EP-2022-001",
    eprint         = "2202.12988",
    archivePrefix  = "arXiv",
    primaryClass   = "hep-ex",
}

@Article{CMS-B2G-21-003,
    author         = "{CMS Collaboration}",
    title          = "{Search for a massive scalar resonance decaying to a light scalar and a Higgs boson in the four \(b\) quarks final state with boosted topology}",
    journal        = "Phys. Lett. B",
    volume         = "842",
    year           = "2023",
    pages          = "137392",
    doi            = "10.1016/j.physletb.2022.137392",
    reportNumber   = "CERN-EP-2022-034",
    eprint         = "2204.12413",
    archivePrefix  = "arXiv",
    primaryClass   = "hep-ex",
}

@Article{CMS-HIG-21-008,
    author         = "{CMS Collaboration}",
    title          = "{Search for Higgs Boson Decay to a Charm Quark-Antiquark Pair in Proton--Proton Collisions at \(\sqrt{s} = 13\,\text{TeV}\)}",
    journal        = "Phys. Rev. Lett.",
    volume         = "131",
    year           = "2023",
    pages          = "061801",
    doi            = "10.1103/PhysRevLett.131.061801",
    reportNumber   = "CERN-EP-2022-081",
    eprint         = "2205.05550",
    archivePrefix  = "arXiv",
    primaryClass   = "hep-ex",
}

@Article{CMS-TOP-21-014,
    author         = "{CMS Collaboration}",
    title          = "{Measurement of the \(t\bar{t}\) charge asymmetry in events with highly Lorentz-boosted top quarks in \(pp\) collisions at \(\sqrt{s} = 13\,\text{TeV}\)}",
    journal        = "Phys. Lett. B",
    volume         = "846",
    year           = "2023",
    pages          = "137703",
    doi            = "10.1016/j.physletb.2023.137703",
    reportNumber   = "CERN-EP-2022-155",
    eprint         = "2208.02751",
    archivePrefix  = "arXiv",
    primaryClass   = "hep-ex",
}

@Article{CMS-B2G-22-003,
    author         = "{CMS Collaboration}",
    title          = "{Search for Nonresonant Pair Production of Highly Energetic Higgs Bosons Decaying to Bottom Quarks}",
    journal        = "Phys. Rev. Lett.",
    volume         = "131",
    year           = "2023",
    pages          = "041803",
    doi            = "10.1103/PhysRevLett.131.041803",
    reportNumber   = "CERN-EP-2022-090",
    eprint         = "2205.06667",
    archivePrefix  = "arXiv",
    primaryClass   = "hep-ex",
}

@Booklet{ATL-PHYS-PUB-2014-021,
    author         = "{ATLAS Collaboration}",
    title          = "{ATLAS Pythia~8 tunes to \(7~\text{TeV}\) data}",
    howpublished   = "{ATL-PHYS-PUB-2014-021}",
    url            = "https://cds.cern.ch/record/1966419",
    year           = "2014",
}

@Booklet{ATL-PHYS-PUB-2015-051,
    author         = "{ATLAS Collaboration}",
    title          = "{Early Inner Detector Tracking Performance in the 2015 Data at \(\sqrt{s} = 13~\text{TeV}\)}",
    howpublished   = "{ATL-PHYS-PUB-2015-051}",
    url            = "https://cds.cern.ch/record/2110140",
    year           = "2015",
}

@Booklet{ATL-PHYS-PUB-2016-020,
    author         = "{ATLAS Collaboration}",
    title          = "{Studies on top-quark Monte Carlo modelling for Top2016}",
    howpublished   = "{ATL-PHYS-PUB-2016-020}",
    url            = "https://cds.cern.ch/record/2216168",
    year           = "2016",
}

@Booklet{ATL-PHYS-PUB-2017-015,
    author         = "{ATLAS Collaboration}",
    title          = "{Improving jet substructure performance in ATLAS using Track-CaloClusters}",
    howpublished   = "{ATL-PHYS-PUB-2017-015}",
    url            = "https://cds.cern.ch/record/2275636",
    year           = "2017",
}

@Booklet{ATL-PHYS-PUB-2021-011,
    author         = "{ATLAS Collaboration}",
    title          = "{Evaluating statistical uncertainties and correlations using the bootstrap method}",
    howpublished   = "{ATL-PHYS-PUB-2021-011}",
    url            = "https://cds.cern.ch/record/2759945",
    year           = "2021",
}

@Booklet{ATL-PHYS-PUB-2021-028,
    author         = "{ATLAS Collaboration}",
    title          = "{Identification of hadronically-decaying top quarks using UFO jets with ATLAS in Run~2}",
    howpublished   = "{ATL-PHYS-PUB-2021-028}",
    url            = "https://cds.cern.ch/record/2776782",
    year           = "2021",
}

@Booklet{ATL-PHYS-PUB-2022-039,
    author         = "{ATLAS Collaboration}",
    title          = "{Constituent-Based Top-Quark Tagging with the ATLAS Detector}",
    howpublished   = "{ATL-PHYS-PUB-2022-039}",
    url            = "https://cds.cern.ch/record/2825328",
    year           = "2022",
}

@Booklet{ATL-PHYS-PUB-2023-017,
    author         = "{ATLAS Collaboration}",
    title          = "{Tagging boosted \(W\) bosons applying machine learning to the Lund Jet Plane}",
    howpublished   = "{ATL-PHYS-PUB-2023-017}",
    url            = "https://cds.cern.ch/record/2864131",
    year           = "2023",
}

@Booklet{ATL-PHYS-PUB-2023-020,
    author         = "{ATLAS Collaboration}",
    title          = "{Constituent-Based \(W\)-boson Tagging with the ATLAS Detector}",
    howpublished   = "{ATL-PHYS-PUB-2023-020}",
    url            = "https://cds.cern.ch/record/2860189",
    year           = "2023",
}

@Booklet{ATL-PHYS-PUB-2023-021,
    author         = "{ATLAS Collaboration}",
    title          = "{Transformer Neural Networks for Identifying Boosted Higgs Bosons decaying into \(b\bar{b}\) and \(c\bar{c}\) in ATLAS}",
    howpublished   = "{ATL-PHYS-PUB-2023-021}",
    url            = "https://cds.cern.ch/record/2866601",
    year           = "2023",
}

@Booklet{ATL-PHYS-PUB-2023-032,
    author         = "{ATLAS Collaboration}",
    title          = "{Constituent-Based Quark Gluon Tagging using Transformers with the ATLAS detector}",
    howpublished   = "{ATL-PHYS-PUB-2023-032}",
    url            = "https://cds.cern.ch/record/2878932",
    year           = "2023",
}

\FloatBarrier

\clearpage
 
\begin{flushleft}
\hypersetup{urlcolor=black}
{\Large The ATLAS Collaboration}

\bigskip

\AtlasOrcid[0000-0002-6665-4934]{G.~Aad}$^\textrm{\scriptsize 104}$,
\AtlasOrcid[0000-0001-7616-1554]{E.~Aakvaag}$^\textrm{\scriptsize 17}$,
\AtlasOrcid[0000-0002-5888-2734]{B.~Abbott}$^\textrm{\scriptsize 123}$,
\AtlasOrcid[0000-0002-0287-5869]{S.~Abdelhameed}$^\textrm{\scriptsize 119a}$,
\AtlasOrcid[0000-0002-1002-1652]{K.~Abeling}$^\textrm{\scriptsize 56}$,
\AtlasOrcid[0000-0001-5763-2760]{N.J.~Abicht}$^\textrm{\scriptsize 50}$,
\AtlasOrcid[0000-0002-8496-9294]{S.H.~Abidi}$^\textrm{\scriptsize 30}$,
\AtlasOrcid[0009-0003-6578-220X]{M.~Aboelela}$^\textrm{\scriptsize 45}$,
\AtlasOrcid[0000-0002-9987-2292]{A.~Aboulhorma}$^\textrm{\scriptsize 36e}$,
\AtlasOrcid[0000-0001-5329-6640]{H.~Abramowicz}$^\textrm{\scriptsize 154}$
% ...
\bigskip
\\

% ...
$^{17}$Department for Physics and Technology, University of Bergen, Bergen; Norway.\\
% ...
$^{30}$Physics Department, Brookhaven National Laboratory, Upton NY; United States of America.\\
% ...
$^{36}$$^{(a)}$Facult\'e des Sciences Ain Chock, Universit\'e Hassan II de Casablanca;$^{(b)}$Facult\'{e} des Sciences, Universit\'{e} Ibn-Tofail, K\'{e}nitra;$^{(c)}$Facult\'e des Sciences Semlalia, Universit\'e Cadi Ayyad, LPHEA-Marrakech;$^{(d)}$LPMR, Facult\'e des Sciences, Universit\'e Mohamed Premier, Oujda;$^{(e)}$Facult\'e des sciences, Universit\'e Mohammed V, Rabat;$^{(f)}$Institute of Applied Physics, Mohammed VI Polytechnic University, Ben Guerir; Morocco.\\
% ...
$^{45}$Physics Department, Southern Methodist University, Dallas TX; United States of America.\\
% ...
$^{50}$Fakult\"{a}t Physik , Technische Universit{\"a}t Dortmund, Dortmund; Germany.\\
% ...
$^{56}$II. Physikalisches Institut, Georg-August-Universit\"{a}t G\"ottingen, G\"ottingen; Germany.\\
% ...
$^{104}$CPPM, Aix-Marseille Universit\'e, CNRS/IN2P3, Marseille; France.\\
% ...
$^{119}$$^{(a)}$New York University Abu Dhabi, Abu Dhabi;$^{(b)}$United Arab Emirates University, Al Ain; United Arab Emirates.\\
% ...
$^{123}$Homer L. Dodge Department of Physics and Astronomy, University of Oklahoma, Norman OK; United States of America.\\
% ...
$^{154}$Raymond and Beverly Sackler School of Physics and Astronomy, Tel Aviv University, Tel Aviv; Israel.\\
% ...
$^{a}$ Also Affiliated with an institute covered by a cooperation agreement with CERN.\\
$^{b}$ Also at An-Najah National University, Nablus; Palestine.\\
$^{c}$ Also at Borough of Manhattan Community College, City University of New York, New York NY; United States of America.\\
$^{d}$ Also at Center for Interdisciplinary Research and Innovation (CIRI-AUTH), Thessaloniki; Greece.\\
$^{e}$ Also at Centro Studi e Ricerche Enrico Fermi; Italy.\\
$^{f}$ Also at CERN, Geneva; Switzerland.\\
% ...

\end{flushleft}


%

%
%

%

%
%
%
%

\end{document}